\documentclass[prd,aps,showpacs,frontmatterverbose,amsmath,amsymb,twocolumn,groupedaddress]{revtex4-1}

\usepackage{color}

\newcommand{\grad}{\mbox{\boldmath{$\nabla$}}}

\newcommand{\MyBox}{\Box}
\newcommand{\vdot}{\cdot}
\newcommand{\wave}{\Phi}

\newcommand{\pie}{\mathfrak{p}}
\newcommand{\h}{\mathfrak{h}}
\newcommand{\const}{\mathfrak{c}}
\newcommand{\T}{\mathfrak{T}}
\newcommand{\F}{\mathfrak{F}}
\newcommand{\vecs}[1]{\mathbf{#1}}

\newcommand{\pha}{\varphi}
\newcommand{\Ricci}{ {}^3 R
 }
\newcommand{\extra}{\chi_1}
\newcommand{\extratwo}{\chi_2}

\usepackage{graphicx}
\usepackage{dcolumn}
\usepackage{bm}
\usepackage{amsmath}
\usepackage{amssymb}

\def\eg{\hbox{e.g.}\ }
\def\ie{\hbox{i.e.}\ }

\begin{document}


\title{Canonical reduction for dilatonic gravity in $3+1$ dimensions}

\author{T. C. Scott}
\email{tcscott@gmail.com}
\affiliation{
College of Physics and Optoelectronics, Taiyuan University of Technology, Shanxi 030024, China \\
}
\affiliation{
Near India Pvt Ltd, no. 71/72, Jyoti Nivas College Road, Koramangala, Bengalore 560095, India\\
}

\author{Xiangdong Zhang}
\email{Corresponding author.\\ scxdzhang@scut.edu.cn}
\affiliation{
Department of Physics, South China University of Technology, Guangzhou 510641, China\\
}
\author{R. B. Mann}
 \email{rbmann@uwaterloo.ca}
\affiliation{Physics Department, University of Waterloo, Waterloo, Ontario, N2L-3G1, Canada\\
}
\affiliation{
Perimeter Institute, 31 Caroline St. N. Waterloo Ontario, N2L 2Y5, Canada\\
}
\author{G. J. Fee}
\email{gjfee@cecm.sfu.ca}
\affiliation{Centre for Experimental and Constructive Mathematics (CECM),
    Simon Fraser University, Burnaby, British Columbia V5A 1S6, Canada\\
}
\homepage[Originally published in Physical Review D {\bf 93}, 084017 (2016), the
present version has two additional appendices.  In particular Appendix \ref{appa} has been extended into a more self-contained derivation.]{}

\date{\today}
\begin{abstract}
We generalize the $1+1$-dimensional gravity formalism of Ohta and Mann
to $3+1$ di\-men\-sions by developing the canonical reduction of a proposed
formalism applied to a system coupled with a set of point particles. This
is done via the Arnowitt-Deser-Misner method and
by eliminating the resulting constraints and imposing coordinate conditions.
The reduced Hamiltonian is completely determined in terms of
the particles' canonical variables (coordinates, dilaton field and momenta).
It is found that the equation governing the dilaton field under
suitable gauge and coordinate conditions, including the absence of transverse-traceless
metric components, is a {\em logarithmic Schr\"{o}dinger equation}.
Thus, although different, the
$3+1$ formalism retains some essential features of the earlier $1+1$ formalism,
in particular the means of obtaining a quantum theory for dilatonic gravity.
\end{abstract}

\maketitle


\section{Introduction}


Two of the oldest and most notoriously vexing prob\-lems in gravitational theory (which are possibly related to each other) are $ \rm (i)$
obtaining a quantum gravity theory which is renormalizable and therefore amenable to meaningful physical predictions, and $ \rm (ii)$
determining the (self-consistent) motion of $N$ bodies and the resultant metric they collectively
produce under their mu\-tual gra\-vi\-ta\-tional 
in\-fluence\cite{mann:gravity1}. In the latter case, lower-dimensional theories such as
 $1+1$ dimensional gravity,
(meaning one spatial dimension and one time dimension) have been examined
 in large part because problems in quantum gravity
become much more mathematically tractable in this context.  
However, the problematic
issue for General Relativity (GRT) is that the Einstein tensor
is topologically trivial in $1+1$ dimensions and cannot yield the correct Newtonian
limit.  Through the addition of an auxiliary field corresponding
to a particle known as a {\em dilaton}, this problem can be addressed and yields
a successful many-body gravity theory \cite{2dross,mann:gravity0, mann:gravity1}.

Regarding the first issue, in lower di\-men\-sions, namely $1+1$, a normalizable
quantum theory combining gravity, quantum mechanics and even an electromagnetic
interaction was found through the addition of a dilaton\cite{mann:gravity2}.
It also reduces to the Newtonian $N$-body gravitational action in the nonrelativistic limit \cite{Mann:2001fg,Burnell:2002ps,Burnell:2003iu}. 
The action for the gravitational scalar-tensor formulation in $1+1$ di\-men\-sions \cite{mann:gravity3,mann:gravity0,mann:gravity1,mann:gravity2}
 coupled to $N$ particles is,  in the presence of a cosmological constant
$\Lambda$ 
\begin{widetext}
\begin{equation}
I=\int d^{2}x\left[
\frac{1}{2\kappa}\sqrt{-g}g^{\mu\nu}
\left\{\Psi R_{\mu\nu}+\frac{1}{2}\nabla_{\mu}\Psi\nabla_{\nu}\Psi
+\frac{1}{2}g_{\mu\nu}\Lambda \right\}
-
\sum_{a} m_{a}\int d\tau_{a}
\left\{-g_{\mu\nu}(x)\frac{dz^{\mu}_{a}}{d\tau_{a}}
\frac{dz^{\nu}_{a}}{d\tau_{a}}\right\}^{1/2}\delta^{2}(x-z_{a}(\tau_{a}))
\right] \label{e1} \\
\end{equation}
\end{widetext}
where $\Psi$ is the  auxiliary scalar field. Here $g_{\mu\nu}$,
$g$, $R$ and $\tau_{a}$ are the metric tensor of space\-time,
det$(g_{\mu\nu})$,
the Ricci scalar and the proper time of the $a$ th particle, respectively,
and $\kappa=8\pi G/c^4$. The symbol $\grad_{\mu}$ denotes the covariant
derivative associated with $g_{\mu\nu}$.
 
 The field equations derived from the
variations $\delta g_{\mu\nu}$ and $\delta\Psi$ are
\begin{eqnarray} \label{e4}
&& - \grad_{\mu} \grad_{\nu} \Psi + g_{\mu \nu } ( \MyBox \Psi - \frac{1}{4} (\grad \Psi )^2  )  \\
&& \qquad \qquad + \frac{1}{2}
 \grad_{\mu} \Psi \grad_{\nu} \Psi  =  \kappa T_{\mu\nu}+\frac{1}{2}g_{\mu\nu}\Lambda  ~, \nonumber
\end{eqnarray}
\begin{equation}
R-g^{\mu\nu}\grad_{\mu}\grad_{\nu}\Psi=R-\frac{1}{\sqrt{-g}}\partial_{\mu}
(\sqrt{-g}g^{\mu\nu}\partial_{\nu}\Psi)=0  \, , \label{eq:R}
\end{equation}
where
\begin{eqnarray}
(\grad \Psi )^2 & = &  g^{\mu \nu } ( \grad_{\mu} \Psi)  \grad_{\nu} \Psi = \grad^{\lambda} \Psi \grad_{\lambda} \Psi ~,  \nonumber \\
\MyBox \Psi & = & g^{\mu \nu } \grad_{\mu} \grad_{\nu} \Psi = \grad^{\lambda} \grad_{\lambda} \Psi \nonumber \\
& = & \frac{1}{\sqrt{-g}}\partial_{\mu}
(\sqrt{-g}g^{\mu\nu}\partial_{\nu}\Psi), \label{eq:box}
\end{eqnarray}
is the d'Alembertian (four-dimensional Laplace-Beltrami operator) and
\begin{eqnarray}
T_{\mu\nu}&=&-\frac{2}{\sqrt{-g}}\frac{\delta\cal L\mit_{M}}{\delta g^{\mu\nu}}  \\
&=&\sum_{a} m_{a}\int d\tau_{a}\frac{1}{\sqrt{-g}}
g_{\mu\sigma}g_{\nu\rho}\frac{dz^{\sigma}_{a}}{d\tau_{a}}
\frac{dz^{\rho}_{a}}{d\tau_{a}}\delta^{2}(x-z_{a}(\tau_{a}))\;\; \nonumber ,
\end{eqnarray}
$\cal L\mit_{M}$ being the matter Lagrangian gi\-ven by the second term
in brackets on the right-hand side of (\ref{e1}).
The trace of Eq.~(\ref{e4}) is
\begin{equation}
\MyBox \Psi = \grad^{\mu}\grad_{\mu}\Psi=\kappa T^{\mu}_{\;\;\mu}+\Lambda \;\;,
\end{equation}
which yields
\begin{equation}\label{RT}
R=\kappa T^{\mu}_{\;\;\mu}+\Lambda \;\;.
\end{equation}
(hence, this is called $R=T$ theory).
If the stress-energy tensor of the particles is absent the above equation reduces to that considered in  earlier work of Jackiw \cite{jackiw1,jackiw2} and Teitelboim \cite{teitelboim}.  Although
GRT yields trivial field equations in
 $1+1$ di\-men\-sions,  incorporating a dilaton in the manner shown in (\ref{e1})
 ensures a nontrivial set of field equations with the correct Newtonian li\-mit \cite{mann:gravity1}.

It was later found that in $1+1$ di\-men\-sions   the above $N$-body problem could be mapped onto the quantum-mechanical problem of an $N$-body generalization of the problem of the $\textrm{H}^+_2$ molecular ion in one dimension,  combining into a normalizable theory represented by the Schr\"{o}dinger equa\-tion \cite{mann:gravity2}. The formalism could also be extended to include electromagnetic charges.  However, since our world is in $3+1$ di\-men\-sions, the impact of this work is not yet clear.

Our proposed generalization of the action \eqref{e1} 
is simply
the outcome of the $d+1$ generalization of
Eq.~(\ref{e1}) as written in Sec.~9 of Ref.~\cite{mann:gravity1} 
\ie{}
$\int d^{n+1} x  {\cal L}$ 
where [Eq.~$(6.1)$ of \cite{adm0}, Eq.~$(2.1)$ of \cite{r7}]
\begin{widetext}
\begin{eqnarray}\label{Lag1}
{\cal L} &=& {\cal L}_F + \cal L\mit_{M} =  \frac{2}{\kappa^{2}}\sqrt{-g}\left\{\Psi R
+\frac{1}{2}g^{\mu\nu}\grad_{\mu}\Psi\grad_{\nu}\Psi - 2 \Lambda \right\}
-\frac{1}{2}\sum_{a}\sqrt{-g}\left(g^{\mu\nu}\pha_{a,\mu}\pha_{a,\nu}
+m^{2}_{a}\pha^{2}_{a}\right) \nonumber\\
&\quad & +\frac{1}{2} (A_{\mu, \nu} -  A_{\nu , \mu} ){\cal F}^{\mu \nu}  + \frac{1}{4} \sqrt{-g} ~ {\cal F}^{\mu \nu } {\cal F}^{\alpha \beta } g_{\mu \alpha } g_{\nu \beta } + \Sigma_{i} \int ds ~ e_i \left(\frac{{dx_i}^{\mu} (s)}{ds} \right) A_{\mu} (x) ~ \delta^4 (x_i -x(s) ) \\
&+& \Sigma_{i} \int ds~ \left\{ {p_i}_{\mu} \left(\frac{{dx_i}^{\mu}(s) }{ds} \right) \right. \label{eq:LM}  - \left. \frac{1}{2} \lambda_i (s)
( {p_i}_{\mu} {p_i}_{\nu} g^{\mu \nu} + m_i^2 ) \right\} ~ \delta^4 (x_i -x(s) ) \nonumber
\end{eqnarray}
\end{widetext}
where $\kappa^{2}=32\pi G/c^4$ (note the redefinition of $\kappa$ compared to the  $1+1$ theory), and where  $ {\cal F}_{\mu \nu } = A_{\mu, \nu} -  A_{\nu , \mu}$ is an electromagnetic field strength tensor density  whose gauge potential is $A_{\mu}$. 
Note that ${p_i}_{\mu}$ is the mechanical momentum, $e_i$ is the charge and $\lambda_i $ is the Lagrange multiplier of the $i${th} particle. We have included coupling to   $N$ neutral massive scalar fields $\pha_a$, which for certain purposes can be used instead of point particles in studying the $N$-body problem.

The reasons for using \eqref{Lag1} are as follows.
As shown in Ref. \cite{mann:gravity1}, the $d+1$ generalization of
Eq.~(\ref{e1}) which includes the dilaton guarantees the correct Newtonian
li\-mit in $d+1$ di\-men\-sions.  This was proven for $d=1,2,3$
in Sec.~9 of \cite{mann:gravity1} and, since it is a vital cornerstone in our proposed
generalization, the proof is reproduced here in Appendix \ref{appa} with more detail.

Note that dimensional scaling pioneered by Hershbach {\em et al.} \cite{dimscal}
has provided much insight in quantum theory and is suggestive of sound theories.
For example, dimensional scaling helped establish
that the mathematical structure of the energy eigenvalues for the three-dimensional hydrogen
molecular ion was a generalized Lambert $W$ function \cite{scott:aaecc} from its simpler
one-dimensional counterpart, the double-well Dirac-delta function model \cite{scott:chemphys}.

An obvious criticism is that in $3+1$ di\-men\-sions, the scalar-tensor theory of Eq.~(\ref{Lag1})
is clearly a departure from GRT.  However, if we let \cite{mann:gravity1} 
\begin{equation}
\Psi= 1+\kappa\psi\; \label{dila}
\end{equation}
where $\kappa$ is small and take the li\-mit $\kappa \rightarrow 0$, then $\Psi \rightarrow 1$ and Eq.~(\ref{Lag1})
reduces to the familiar Einstein-Hilbert action of GRT.  The results for Appendix \ref{appa} prove that both GRT and our
scalar-tensor theory of Eq.~(\ref{Lag1}) yield the same correct Newtonian li\-mit in $3+1$ di\-men\-sions. Thus
in $3+1$ di\-men\-sions, the effect of the dilaton is very small; this is salutary as we do not, {\em a priori}, expect it to contradict experiments
vindicating GRT to known accuracies as given by, e.g., the Gravity Probe A and B experiments
\cite{Will:2014kxa}.

Another reason for retaining the dilaton is the observation of the unusual resemblance reported between the dilaton
(a particle whose origin can be traced to Kaluza-Klein theory) and
the Higgs boson (from the standard model) to the extent that a number of authors have wondered if they represent two  different signatures for the same
particle and so might even be the same particle (\eg see the work of Bellazzini {\em et al.} \cite{higgs} and references therein).
Of course, it will take time for experiments to sort out this issue, but it becomes tantalizing to consider that perhaps the dilaton is closer to being discovered experimentally than the graviton. Thus retaining the dilaton becomes timely and instructive.

This paper is intended as a first in a series to flesh out the proposed $3+1$
scalar-tensor theory 
as a possible foundation for dilatonic
quantum gravity through a canonical reduction of Eq.~(\ref{Lag1}).
The goal of this work is to isolate the effective field equa\-tion governing the dilaton field.  This
is done as follows.
After obtaining the field equa\-tions, we apply the Arnowitt-Deser-Misner (ADM) method \cite{adm0} to our scalar-tensor
theory as it proven useful in decoupling the field equa\-tions for the $1+1$ case.  Next, we then eliminate variables
while trying to retain the greatest generality.  We examine the behavior in the far field and, in general,
under suitable gauge and coordinate conditions.  Finally, we obtain the essential partial differential equa\-tions (PDEs) governing the dilaton field,
the canonical momenta and the metric, and see how the outcome relates
 to that of the $1+1$ case.  Concluding remarks are made at the end.

Throughout the
paper, we use the Greek alphabet for space\-time indices,
the latin alphabet $a; b; c; \ldots$, for spatial indices, and
$i, j, k, \ldots$, for internal or particle indices.


\section{Hamiltonian Analysis}

To obtain the field equa\-tions, we rely on the results of Zhang and Ma (Sec.~II of \cite{xiangdong1}).  Although the latter work was aimed
towards loop quantum gravity (which is not the intent of the current work), the initial derivations of their Hamiltonian analysis use the ADM approach in the context of $f(R)$ gravity and thus their field equa\-tions (obtained before
the injection of the Ashtekar variables in Sec.~III of \cite{xiangdong1}) 
can be extracted (we only use Sec.~ II of their work).

\subsection{Field equations}

In terms of the their own notation, the settings for their coupling parameter
$\omega$ and potential $\xi$ are \cite{xiangdong1} 
\[
 \omega(\Psi) = - \frac{1}{2} \Psi ~, \quad \xi (\Psi) = - \Lambda \quad \mbox{and} \quad 8 \pi G = 1
\]
Variation with respect to $g_{\mu \nu}$ yields 
(in $n=d+1$ spacetime dimensions)
[Eq.~$(2.2)$ of \cite{xiangdong1}]
\begin{eqnarray}
&& \Psi G_{\mu \nu} - \grad_{\mu} \grad_{\nu} \Psi + g_{\mu \nu } \left( \MyBox \Psi - \frac{1}{4} (\grad \Psi )^2 + \Lambda \right)  \label{newe4} \\
&& \quad + \frac{1}{2}
 (\partial_{\mu} \Psi) \partial_{\nu} \Psi  = \frac{\kappa^{2}}{4} T_{\mu \nu } = \frac{8 \pi G}{c^4} T_{\mu \nu } \nonumber
\end{eqnarray}
which has the same functional form of Eq.~(\ref{e4}) apart from a nonzero Einstein tensor $G_{\mu \nu }$ and a slight rescaling
of the definitions for $T_{\mu \nu }$ and the gravitational constant (and of course the realization that a covariant derivative
on a scalar is just the partial derivative $\grad_{\mu} \Psi = \partial_{\mu} \Psi $).  In the li\-mit $\Psi \rightarrow 1$, Eq.~(\ref{newe4})
reduces to the standard Einstein field equa\-tions.
Variation with respect to $\Psi$
yields [Eq.~ $(2.3)$ of \cite{xiangdong1}]
\begin{equation}
R = \MyBox \Psi \label{eq:R3p1}
\end{equation}
as in Eq.~(\ref{eq:R}).  However this is not $R=T$ theory.  Rather the trace of Eq.~(\ref{newe4}) yields, using Eq.(\ref{eq:R3p1})
\begin{equation}
 (n-1) R + (\Psi R + \frac{1}{2} \left( \grad \Psi )^2 \right) \left( 1 - \frac{n}{2} \right) = \frac{\kappa^2}{4} T_{\mu}^{\mu}  - n \Lambda \label{eq:trace}
\end{equation}
When $n=2$, Eq.~(\ref{eq:trace}) does reduce to $R = {\rm const.} \times T_{\mu}^{\mu}$ for $\Lambda = 0$.  However, for $n=4$, it becomes
\begin{equation}
( 3   - \Psi ) R - \frac{1}{2} (\grad \Psi )^2  = \frac{\kappa^2}{4} T_{\mu}^{\mu} - 4 \Lambda  \label{eq:R4}
\end{equation}
Zhang and Ma set $8 \pi G = 1 $ \cite{xiangdong1}.  We will do something very similar and henceforth multiply the Lagrangian of Eq.~(\ref{Lag1}) by
$\kappa^2/2$.

\subsection{ADM method}

Our derivation will follow the general ideas from the original ADM method \cite{adm1,adm2}. In this formalism, the
metric is defined as 
\begin{equation}\label{ADMmet}
ds^2 = -N^2 dt^2 + g_{ab} (dx^a + N^a dt )(dx^b + N^b dt )
\end{equation}
where $N = {\textstyle (-g^{00})^{-1/2} } $ and $N_b = g_{0b}$ are the {\em lapse} function and the {\em shift} covector.
Here $\gamma_{ab} = g_{ab}$ is the $3$-metric for the spatial coordinates of $g_{ab}$ and $\sqrt{\gamma}$ is the square root
of the determinant of $\gamma_{ab}$ where  $\sqrt{-g} = N \sqrt{\gamma}$.
By doing a $3+1$ decomposition of the space\-time, the four-dimensional
scalar curvature can be expressed as [Eq.~ $(2.4)$ of \cite{xiangdong1}] 
\begin{eqnarray}
R & = & K_{ab} K^{ab} - K^2 + \Ricci + \frac{2}{\sqrt{-g}} \partial_{\mu}
( \sqrt{-g} n^{\mu} K ) \nonumber \\
&- & \frac{2}{N \sqrt{\gamma}} \partial_a \left( \sqrt{\gamma} \gamma^{ab} \partial_b N \right) ~,  \label{eq:curv} 
\end{eqnarray}
where $K_{ab}$ is the extrinsic curvature of a spatial hypersurface
$\Sigma$, $K = K_{ab} \gamma^{ab}$, $\Ricci$ denotes the scalar curvature of the
$3$-metric $\gamma_{ab}$ induced on $\Sigma$ and  $n^{\nu}$ is the unit normal of $\Sigma$.
The canonical momenta are respectively [Eq.~$(2.5)$ of \cite{xiangdong1}]
\begin{eqnarray}
\pi^{a b}&=&\frac{\partial {\cal L}} {\partial(\partial_{0} \gamma_{ab})} = \frac{\partial L} {\partial \dot{\gamma_{ab}} } \label{eq:pi} \\
& = & \frac{\sqrt{\gamma}}{2} \left[ \Psi (K^{ab} - K \gamma^{ab} ) - \frac{\gamma^{ab}}{N}
( \dot{\Psi} - N^c \partial_c \Psi ) \right] \nonumber
\end{eqnarray}
and [Eq.~$(2.6)$ of \cite{xiangdong1}]
\begin{equation}
\Pi =\frac{\partial {\cal L}} {\partial(\partial_{0} \Psi)} = \frac{\partial L} {\partial \dot{\Psi} }
 = - \sqrt{\gamma} \left( K + \frac{1}{2N}
( \dot{\Psi} - N^c \partial_c \Psi ) \right) \label{eq:phi}
\end{equation}
where $N^c$ is again the shift vector. Combining the trace of
Eqs.~(\ref{eq:pi}) and (\ref{eq:phi}) gives [Eq.~$(2.7)$ of \cite{xiangdong1}]
\begin{equation}
(3 - \Psi )   ( \dot{\Psi} - N^c \partial_c \Psi ) = \frac{2 N}{\sqrt{\gamma} } (\Psi \Pi - \pi )
\label{eq:ppi}
\end{equation}
where $\pi = \pi_{a b} \gamma^{ab} $.
Note that we can write 
\begin{equation}
n^{\rho} \nabla_{\rho } \Psi = \frac{1}{N} (\dot{\Psi } - N^c \partial_c \Psi ) \label{eq:def}
\end{equation}
using $n^0 = \frac{1}{N}$ and $ n^a = - \frac{N^a}{N}$.
The total Hamiltonian can be derived as a linear combination of constraints as 
\begin{equation}
H_{total} = \int_{\Sigma} d^3 x (N^a V_a + N H )
\end{equation}
where the smeared diffeomorphism and Hamiltonian constraints
read, respectively [Eqs.~$(2.8)$ and $(2.9)$ of \cite{xiangdong1}]
\begin{eqnarray}
V(\vecs{N} ) & = & \int_{\Sigma} d^3 x N^a V_a \label{eq:VN} \\
 &=& \int_{\Sigma} d^3 x N^a \left( -2 D^b (\pi_{ab} ) + \Pi \partial_a \Psi - 
 {\F}_a
 \right) \nonumber
\end{eqnarray}
and
\begin{eqnarray}
H (N) & = & \int_{\Sigma} d^3 x N H \label{eq:H} \\
& = & \int_{\Sigma} d^3 x N \left( \pi^{ab} \dot{H}_{ab} + \Pi \dot{\Phi} - {\cal L} \right) \nonumber \\
& = & \int_{\Sigma} d^3 x N \left[ \frac{2}{\sqrt{\gamma}} \left( \frac{\pi^{ab} \pi_{ab} - \frac{1}{2} \pi^2  }{\Psi}
+ \frac{ (\pi - \Psi \Pi )^2}{\Psi(3-\Psi)} \right) \right. \nonumber \\
& + & \left.  \frac{1}{2} \sqrt{\gamma} \left( - \Ricci \Psi - \frac{1}{2} ( D_a \Psi ) D^a \Psi + 2 D_a D^a \Psi - 2 \Lambda \right) \right.
\nonumber \\
&  - & \left. 
{\F}_0
\right] \nonumber
\end{eqnarray}
where $D_a$ is the covariant
derivative with respect to the $3$-metric $\gamma_{ab} = g_{ab} $.
Note that $D_a \Psi = \partial_a \Psi$ because
$\Psi$ is a scalar
and so [Eq.~$(A.4)$ of \cite{xiangdong1}]
\begin{eqnarray}
( D_a \Psi ) D^a \Psi & = & \gamma^{ab} ( D_a \Psi ) D_b \Psi =  \gamma^{ab} ( \partial_a \Psi ) \partial_b \Psi \nonumber \\
D_a D^a \Psi & = & \frac{1}{\sqrt{\gamma}} D_a ( \sqrt{\gamma} \gamma^{ab} D_b \Psi )  =  \frac{1}{\sqrt{\gamma}} D_a (\sqrt{\gamma} \gamma^{ab} \partial_b \Psi ) \nonumber \\
& = & D_a  (\gamma^{ab} \partial_b  \Psi ) = \gamma^{ab} \left( \partial_a \partial_b \Psi - \Gamma^c_{ab} \partial_c \Psi \right) , ~~ \label{eq:chris}
\end{eqnarray}
where $\Gamma^c_{ab}$ are the Christoffel symbols of the second kind \cite{gravitation} (see Appendix \ref{appf}).

The terms $\F_{\mu}$ are con\-tri\-bu\-tions from the mat\-ter La\-gran\-gian, such as that which appears in \eqref{Lag1}. 
According to the ADM approach (Sec.~6.2 of \cite{adm1}), the contributions to the constraint equations (\ref{eq:VN})  and (\ref{eq:H})  can be obtained via
\begin{equation}
{\F}_0
 = \frac{\partial \cal L\mit_{M}}{\partial N}
\quad 
 {\F}_a = \frac{\partial \cal L\mit_{M}}{\partial N^a} \label{eq:F}
\end{equation}
upon using (\ref{ADMmet}) in the matter Lagrangian
$ \cal L\mit_{M}$. 
Although we formally include these terms, we will not explicitly resolve their effect on the metric.  Rather, the focus of the present work is on the treatment of the free Lagrangian contribution ${\cal L}_F$ [Eqs.~ (\ref{eq:VN}) and (\ref{eq:H}) but without $\F_{\mu}$].

The approach to solving the constraint equations in $1+1$ dimensions was to obtain coordinate conditions that  would set the conjugate momenta $\Pi$ and $\gamma_{ab}$ to fixed numerical
values \cite{mann:gravity1} and inject them into both the shift covector 
constraint equation and the Hamiltonian equation.  However, as
we shall see, in $3+1$ dimensions the conjugate momenta do not generally collapse into fixed
numbers but are functions of the dilaton field $\Psi$.  Furthermore, the 
$1 / \Psi$ and $3 - \Psi$ terms in Eq.~(\ref{eq:H}) render a solution to the constraints somewhat
unwieldy.   We therefore use other
important relationships determined by Zhang and Ma, including relating the curvature $K_{ab}$ to the momentum $\pi_{ab}$ and $\Pi$ {Eq.~$(2.21)$ of \cite{xiangdong1}], \ie
\begin{equation}
K_{ab} = \frac{2 \left( \pi_{ab} - \frac{(2-\Psi ) }{2(3-\Psi )} \pi \gamma_{ab} \right) }{\Psi \sqrt{\gamma}} - \frac{\Pi \gamma_{ab} }{ (3-\Psi ) \sqrt{\gamma} }
\end{equation}
whose trace with respect to the $3$-metric $\gamma_{ab} $ is 
\begin{equation}
K = \frac{2 \pi  \left( 1 - \frac{ n_{\gamma} (2- \Psi ) }{2 (3-\Psi ) } \right) }{ \Psi \sqrt{\gamma} }
- \frac{n_{\gamma} \Pi   }{ (3-\Psi ) \sqrt{\gamma} } \label{eq:Ktrace}
\end{equation}
where $ n_{\gamma} = 3 $ in $3+1$  dimensions.  
 Furthermore 
$\gamma^{ab}$ and $K^{ab}$ are related to each other by 
\begin{eqnarray}
\pi^{ab} \pi_{ab} - \frac{1}{2} \pi^2 &  = & \frac{\gamma}{4} \left[ \Psi^2 (K^{ab} K_{ab} -  K^2 ) \right. \label{eq:pK} \\
& & \quad  \left. -\frac{3}{2} (n^{\mu} \grad_{\mu } \Psi )^2 - 2 \Psi K n^{\mu} \grad_{\mu } \Psi \right] \nonumber
\end{eqnarray}
where we discern the first two terms of Eq.~(\ref{eq:curv}) for $R$.
These relationships allow us to rewrite the constraint
(\ref{eq:H}) in terms of $K_{ab}$ [Eq.~$(2.17)$ of \cite{xiangdong1}] (valid in $3+1$ dimensions),
\begin{eqnarray}
0 & = & \frac{\sqrt{\gamma} \Psi}{2} \left( K_{ab} K^{ab} - K^2 - \Ricci \right) \label{eq:H2} \\
& + &  \frac{\sqrt{\gamma}}{2} \left( 2 D_a D^a \Psi - \frac{1}{2} (D_a \Psi ) D^a \Psi \right) \nonumber \\
& -&
\sqrt{\gamma} ~ \left[ ( n^{\nu} \partial_{\nu} \Psi ) \left(K + \frac{1}{4}  n^{\mu} \partial_{\mu} \Psi  \right)
 +  \Lambda \right] - {\F}_0\nonumber
\end{eqnarray}
upon
using  (\ref{eq:def}). The structural form of Eq.~(\ref{eq:H2}) is
easier to deal with than Eq.~(\ref{eq:H}) since the problematic
$3-\Psi$ term of Eq.~(\ref{eq:H}) is embedded into the curvature $K_{ab}$.
It is now a matter of eliminating
variables where possible to isolate the equa\-tion governing the dilaton field under general conditions.
Equation (\ref{eq:H2}) will prove most useful in this regard. The remainder of this work 
serves to eliminate many of the variables of the formulations of Ref.~\cite{xiangdong1} to obtain the
final equation governing the dilaton field.

\section{Elimination of Variables}
\label{sec:elim}

\subsection{Elimination of $N$ and $N^a$}

An important simplification results from setting \cite{dewitt} 
\begin{equation}
N = 1 \quad   N_a =0   \label{simp}
\end{equation}
which is allowed at the cost of abandoning 
explicit four-dimensional general covariance
\cite{dirac,natalia1,natalia2}.
The time derivatives of $N$ and $N_a$ are also taken as zero.
These settings are often used when applying standard ADM to GRT.
and are justified with more detail in Appendix \ref{appc}.
This section includes any res\-tric\-tions to the class of solutions from these settings.
Though dif\-fer\-ent, these settings none\-theless agree respectively with Eqs.~$(94)$ and $(95)$ of Ref.~\cite{mann:gravity0} of the values
for a gauge choice in the li\-mit $\kappa \rightarrow 0 $ and approximately agree with the results $N \ne 0$ and $N_a = 0$
for a different gauge choice in the $1+1$ case [Eq.~$(99)$ of \cite{mann:gravity0}].
Note that the canonical theory of GRT does not
directly determine the $N^a$ (Sec.~4 of \cite{r7}); the latter are obtained later by the time evolution of the system, 
through e.g. [Eq.~$(16)$ of \cite{xiangdong1}]
\begin{equation}
\dot{\gamma}_{ab} ~ = ~ 2 N K_{ab} + D_a N_b + D_b N_a \label{eq:gammadot}
\end{equation}
and the consistency of the coordinate and gravitational equations.
Equation (\ref{eq:gammadot}) is nothing other than the definition of the extrinsic curvature $K_{ab}$.
Here, for the $3+1$ case, $n^{\rho} \partial_{\rho } \Psi = \frac{1}{N} (\dot{\Psi } - N^c \partial_c \Psi ) = \dot{\Psi} $.
The d'Alembertian of Eq.~(\ref{eq:box}) reduces to 
\begin{equation}
\MyBox \Psi  = - ( \ddot{\Psi} +  \const ~ \dot{\Psi}  )
+ D_a D^a \Psi ~ .
\label{eq:redbox}
\end{equation}
where
\[
\const  ~ = ~ \partial_t (\ln \sqrt{\gamma} ) 
\]
when $\partial_a N^a  = 0$.
Note that under our choice of coordinate conditions, the coefficient of $\dot{\Psi}$ in Eq.~(\ref{eq:redbox})
will  add nothing to the chosen class of solutions for the dilaton field.

\subsection{Curvature $K$ }
Another important equa\-tion is the left-hand side of Eq.~$(2.15)$ 
in Ref.~\cite{xiangdong1} 
\begin{equation}
\dot{\Pi}  - \partial_a (N^a \Pi ) - \partial_{\mu} (\sqrt{-g} n^{\mu} K )
= - \frac{1}{2}\partial_{\nu} \left( \sqrt{-g}  n^{\nu} n^{\sigma}
\partial_{\sigma} \Psi \right)
\end{equation}
which in light of (\ref{simp}) becomes
\begin{equation}
\frac{\partial}{\partial t} \left[ \Pi - K \sqrt{\gamma} \right]
 = - \frac{1}{2}  \partial_t (\sqrt{\gamma} \dot{\Psi}   )
 \label{eq:Kdot1}
\end{equation}
This also assumes that the spatial derivatives of $N$ and $N^a$ are also zero
(note that the result obtained here will be reiterated further in Sec.~ \ref{sec:coor}).
The combination of Eqs.~(\ref{eq:ppi}), (\ref{eq:Ktrace}) and  (\ref{simp})  yields an expression for $\Pi$,
\begin{equation}
\Pi = -\sqrt{\gamma} K - \frac{1}{2} \sqrt{\gamma}
 \dot{\Psi} \label{Kppi}
\end{equation}
and implicit differentiation of Eq.~(\ref{Kppi}) above with respect to the time $t$  also
yields a result for $\partial_t \Pi$ and is consistent with Eq.~(\ref{eq:Kdot1}) if
\begin{equation}
\partial_t\left(K\sqrt{\gamma}\right)  = 0 \quad \Rightarrow \quad
\dot{K} ~ = ~ - \frac{K}{2} ~ \partial_t \ln (\gamma )
\label{eq:KKdot}
\end{equation}
The expression  above will be examined further but under the assumptions made, we can
see that $K$ is a constant of the motion ( $\dot{K}=0$) if   $\dot{\gamma}$ is zero.

\subsection{Coordinate Conditions}
\label{sec:coor}
We begin by noting that 
any gi\-ven symmetric second rank tensor $f_{ab}$ has the orthogonal decomposition
[Eqs.~$(4.7a)$ of Ref.~\cite{adm0}, Eqs.~ $(2.10)-(2.12)$ of \cite{r8}]
\[f_{ab}=f^{TT}_{ab}+f^{T}_{ab}+f_{a,b}+f_{b,a}\]
where
\begin{eqnarray}
f^{T}_{\;ab}&=&\frac{1}{2}\left(f^{T}\delta_{ab}
-\frac{1}{\triangle}f^{T}_{\;,ab}\right)
\nonumber \\
f^{T}&=&f_{aa}-\frac{1}{\triangle}f_{ab,ab}
\nonumber \\
f_{a}&=&\frac{1}{\triangle}\left(f_{ab,b}-\frac{1}{2\triangle}f_{bc,bca}\right)
\nonumber
\end{eqnarray}
and $f_{ab}^{TT} = f^{T}_{\;ab}-\frac{1}{3}f^{T}\delta_{ab}$ is the
transverse-traceless $(TT)$ part of $f_{ab}$ and $\triangle$ is the
Laplacian for the $3$-metric. We apply this
to $g_{ab}$ and $\gamma^{ab}$. Next, we define 
\begin{eqnarray}
h_{ab} & \equiv & g_{ab}-\delta_{ab} =  \gamma_{ab}-\delta_{ab}\label{star} \\
\pi^{ab} & \rightarrow &   {\pi^{ab}}_{GRT} ~, \nonumber
\end{eqnarray}
and then make an orthogonal decomposition 
\begin{eqnarray}
h_{ab} & = & h^{TT}_{ab}+h^{T}_{ab}+h_{a,b}+h_{b,a} \nonumber \\
\pi^{ab} & = & \pi^{abTT}+\pi^{abT}+\pi^{a}_{\;,b}+\pi^{b}_{\;,a} \label{eq:decomp}
\end{eqnarray}
The definition of $h_{ab}$ in Eq.~(\ref{star}) is that of Kimura 
[Eq.~$(2.4a)$ of \cite{r7}] and especially Ohta 
[Eqs.~$(2.10)-(2.19)$ of \cite{r8}], and not that of ADM.
This is essential for our discussion.
Also bear in mind that $\pi^{ab}$ in Eqs.~(\ref{eq:decomp}) is the GRT quantity.
The coordinate conditions and the
generator are worked out in Appendix \ref{appd}. Equation (\ref{eq:pi}) can be rewritten as
\begin{equation}
\pi^{ab} = -\frac{1}{2} \Psi {\pi^{ab}}_{GRT}  - \frac{\sqrt{\gamma} \gamma^{ab}}{2N} (\dot{\Psi} - N^c \partial_c \Psi ) \label{eq:div}
\end{equation}
where $ {\pi^{ab}}_{\rm GRT} = - \sqrt{\gamma} (K^{ab} - K g^{ij} )$ is the standard definition in ADM [Eq.~$(3.3)$ of \cite{adm0}].
This suggests treating $ {\pi^{ab}}_{\rm GRT}$ as a function
of the coordinates only (for a given time, as is usually the most desirable scenario in standard ADM
applied to GRT) and recasting our scalar-tensor theory into the mold of standard ADM (or nearly so).  To this end, the ADM generator $G$ is developed as shown in Appendix \ref{appd} together with the orthogonal decomposition of Eqs.~(\ref{eq:decomp}) applied to ${\pi^{ab}}_{\rm GRT}$.  This yields
the following coordinate conditions for $ \pi^{ab}$  and $g_{ab}$ 
\begin{eqnarray}
\gamma_{ab} = g_{ab} & = &  \delta_{ab} ( 1 + \frac{1}{2} h^{T}  ) +  h_{ab}^{TT}  \label{eq:cond1} \\
\pi^{aa} & = &  - \frac{\sqrt{\gamma} \gamma^{aa}}{2 N} \dot{\Psi}
\quad (\Psi \ne 0 ) \label{eq:cond2}
\end{eqnarray}
since $N^c = 0 $.
This reduces to the standard result $\pi^{aa}= 0$ in the GRT limit as  $\Psi \rightarrow 1$ as ex\-pec\-ted, or simply if $\dot{\Psi}=0$.
Equation (\ref{eq:cond1}) is the familiar result obtained by standard 
ADM applied to GRT.
In the absence of gra\-vi\-tons, or 
ge\-ne\-rally for $g_{ab}^{TT} = 0$, the
metric $g_{ab}$ reduces to the {\em isotropic} form 
[Eq.~$(4.7)$ of \cite{adm0}], \ie{},
\begin{equation}
g_{ab} = \gamma_{ab} = \delta_{ab} \h = \delta_{ab} (1 + {\textstyle \frac{1}{2}} h^T )  \quad \rightarrow \quad  \sqrt{\gamma} = \h^{3/2}
\label{eq:isotropic}
\end{equation}
with the far-field boundary condition 
\[
\lim_{r \rightarrow \infty} h^T (r) ~ = ~ 0.
\]
In isotropic coordinates $\delta_{ab}\gamma^{ab} = 3 /\h $ and, from Eq.~(\ref{eq:cond2}) 
\begin{eqnarray}
\pi & = & \h ~ \delta_{ab}\pi^{ab} = -\frac{3}{2} ~\frac{\sqrt{ \gamma}}
{N} \dot{\Psi} =  -\frac{3}{2}  \sqrt{ \gamma}  \dot{\Psi} ~
\nonumber \\
\Rightarrow K & = & 0   \label{eq:piiso} \\
\Pi & = & -\frac{1}{2} \sqrt{\gamma} \dot{\Psi} = \frac{1}{3}  \pi  \nonumber
\end{eqnarray}
Substituting $\Pi$ from Eq.~(\ref{eq:piiso}) into an implicit differentiation of Eq.~(\ref{eq:ppi}) with respect to the time $t$ yields 
the same equation, Eq.~(\ref{eq:KKdot}), relating $\dot{K}$ to $K$ and thus 
vindicating it [this is because this recent derivation
did not require explicit assumptions about the spatial derivatives
of $N$ and $N^a$ being zero, but resulted rather from Eq.~(\ref{simp}) and the coordinate conditions].

Thus the simplifications of Eq.~(\ref{simp}) with Eqs.~(\ref{eq:KKdot}) and (\ref{eq:piiso})
lead to $\dot{K} = K = 0$.
Consequently,
the Ricci scalar in Eq.~(\ref{eq:curv}) with (\ref{eq:R3p1}) simplifies to
\begin{equation}
R  = K_{ab} K^{ab} - K^2 + \Ricci  = \MyBox \Psi ~.  
\label{eq:curv2}
\end{equation}
Note that there remains the term $\partial_a \left( \sqrt{\gamma} \gamma^{ab} \partial_b N \right) $ on the far-right side Eq.~(\ref{eq:curv})
but even if $\partial_b N $ is not taken as zero, it can be ``absorbed'' in the Ricci scalar term $ \Ricci$ and added on as discussed
later.
Moreover, if $\Psi$ is time independent, then $\dot{\Psi}$, $\pi$, and $\Pi$ are all zero. As mentioned in Appendix \ref{appa}, the effect
of the graviton can be treated as the $(TT)$ part of the metric and can be handled thanks to \eg Eq.~$(1.55)$ of \cite{hamber},
\[
\sqrt{-g} = \sqrt{-\eta} \left[ 1 + \frac{1}{2} h_{\mu}^{\mu} + \frac{1}{8}  h_{\mu}^{\mu} h_{\nu}^{\nu} - \frac{1}{4}  h_{\mu}^{\nu}  h_{\nu}^{\mu} \ldots \right]
\]
which is valid for any perturbation  $g_{\mu \nu} = \eta_{\mu \nu} + h_{\mu \nu} $, where the rai\-sing and lowering of indices is done with respect to 
any arbitrary background with metric $\eta$.
A $(TT)$ perturbation would consequently only affect $\sqrt{-g}$ at second order. In a number of cases, the $(TT)$ contribution is not used
(\eg{} Kimura's treatment of the two-body problem within the Einstein-Infeld-Hoffman approximation [P.~159 of \cite{r7}]).  
Therefore, in what follows we work  mostly with isotropic coordinates in which  the $(TT)$ contribution 
vanishes to leading order perturbatively.

In general, from the generator of Eq.~(\ref{eq:G}), the action can be rewritten in the tradition of ADM [Eq.~$(4.17)$ of \cite{adm0}],
\begin{equation}
G=G_M ~ + ~ \int dx^3 \left\{ \pi^{abTT} \delta h_{ab}^{TT}
+{\cal T}^0_{\mu} \delta x^{\mu}\right\}
\end{equation}
where $G_M$ is the generator from the matter Lagrangian. Therefore 
\begin{eqnarray}
{\cal T}^0_0 &=& \cal H\mit =\triangle h^T \label{eq:energy}
\\
{\cal T}^0_a &=&-2\partial_b \pi^{ab} \;\;.
\end{eqnarray}
and the Hamiltonian density is in terms of the metric just as in standard GRT, and not $\triangle \Psi$
as it would be in the $1+1$ case.  In the latter, the dilaton field represents most of the dynamics (because conventional GRT yields nothing in $1+1$)
whereas in $3+1$ di\-men\-sions, the dilaton
field is treated as a small departure from GRT. (Appendix \ref{appa} shows that conventional GRT is sufficient to ensure the correct
nonrelativistic limit).
Usually such simplifications would allow us to address the constraint equa\-tions ~(\ref{eq:VN}) and (\ref{eq:H}) in terms of the two
unknowns $\pi^{ij}$ and $h^T$ (related to the metric).  However, we have additionally the dilaton field $\Psi$. This is addressed in the
next sections.

\section{Shift-covector constraint Equation}
\label{sec:covector}
Here, we highlight the possibilities of reduction and simplification to make Eq.~(\ref{eq:VN}) more solvable.
Given Eq.~(\ref{eq:div}), we consider dividing $\pi^{ab}$ into two parts, where the second part depends explicitly on
$\dot{\Psi}$; expressing the momentum in lower indices,
\begin{equation}
\pi_{ab} =  -\frac{1}{2} \Psi {\pi_{ab}}_{GRT} + \pie_{ab}~~~ \mbox{where} ~~~ \pie_{ab} = - \frac{\sqrt{\gamma} \gamma_{ab}}{2} \dot{\Psi}  \label{eq:div2}
\end{equation}
which allows us to simplify the left side of the shift-vector
equation with
\[
D^b (\pie_{ab} ) = - \frac{\sqrt{\gamma} \gamma_{ab}}{2} D^b (\dot{\Psi}) =  - \frac{\sqrt{\gamma}}{2} D_a (\dot{\Psi}) =
- \frac{\sqrt{\gamma}}{2} \partial_a \dot{\Psi}
\]
From Eq.~(\ref{eq:ppi})
\begin{equation}
\Pi \partial_a \Psi ~ = ~ \frac{\sqrt{\gamma}}{2 \Psi} (3-\Psi )\dot{\Psi} \partial_a \Psi
+ \pi \underbrace{\frac{\partial_a \Psi}{\Psi}}_{\partial_a \ln \Psi }  \label{eq:pidiff}
\end{equation}
We see that if $\dot{\Psi}=0$ then both $\pi=0$ [from Eq.~(\ref{eq:cond2})] and
$\pie_{a b} = 0 $ [from Eq.~(\ref{eq:div2})]. Consequently, $\Pi \, \partial_a \Psi  = 0$ in Eq.~(\ref{eq:pidiff}). In such a case,
the shift-vector constraint  (\ref{eq:VN}) reduces to that of
standard GRT and $\pi^{ab}$ can be
readily calculated by existing methods. For a nonzero $\dot{\Psi}$, Appendix \ref{appe} shows that for these particular cases of separability for $\Psi$, \ie
\begin{eqnarray}
\Psi & = & F(x) ~ G(t)  \qquad \mbox{product} \label{sep1} \\
\Psi & = & F(x) + G(t)  \qquad \mbox{sum} \label{sep2}
\end{eqnarray}
the first term on the right-side of Eq.~(\ref{eq:pidiff}) is a {\em divergence}; \ie it yields a vanishing surface term to the integral of the
shift-vector constraint equation over spatial coordinates. Consequently it
does not contribute to the Euler-Lagrange equa\-tions and can therefore be discarded.
In the appendix, we make use of the ``densitized'' lapse function or ``Taub function'', which was introduced by York as a means of improving the ADM approach [Eq.~ $(41)$ of \cite{york1}},
\begin{equation}
\alpha \equiv \frac{N}{\sqrt{\gamma}} \label{eq:taub}
\end{equation}
and which often appears in the context of setting an initial va\-lue problem.  In ADM,
the lapse function tells how the proper time moves a\-long from spa\-tial $\mbox{slice}$ to spa\-tial $\mbox{slice}$ as the coor\-di\-nate time moves.  Its
setting is a matter of choice, and is consequently an additional coordinate (``gauge'') freedom which does {\em not} change the physical solution,
but will change how well posed the problem is \cite{taub1,taub2,taub3} (and what happens if the initial data conditions are slightly
perturbed).   The Taub function appears
in boundary-va\-lue problems \cite{Sarbach:2012pr} and for stabilizing numerical relativity \cite{Calabrese:2002ei,Calabrese:2002ej}.
In our case, clearly $\alpha \rightarrow 1$ in the far-field limit.

Thus for $\dot{\Psi} \ne 0$ and for
the separable cases of Eqs.~(\ref{sep1}) or (\ref{sep2}), the first term of Eq.~(\ref{eq:pidiff}) can be discarded, leaving only
the term proportional  $\pi \partial_a \ln (\Psi ) $, with $\pi$ gi\-ven by Eq.~(\ref{eq:piiso}). In isotropic coordinates this is
also a divergence  and can therefore be discarded, as explained in Appendix \ref{appe}. 

These particular separable solutions of $\Psi$ therefore eliminate 
the $\pi \partial_a \ln (\Psi )$ term, and bring the vector equation of Eq.~(\ref{eq:VN}) much closer to the GRT result.  Consequently we can make use of
existing GRT results to obtain solutions to the vector constraint  
Eq.~(\ref{eq:VN}).

We now focus exclusively on the first term of Eq.~(\ref{eq:VN}) and since the indices of individual components can be raised and lowered
with the metric, \eg  $D^b (\pi_{ab} ) = D_b (\pi^{b}_a ) $, 
we make use of an important result in ADM [Sec.~ 3 of \cite{adm2}],
which allows us to convert the covariant derivative into a simple partial derivative, \ie
\begin{equation}
D^b (\pi_{ab} ) ~ \rightarrow  ~ \partial_b  \pi^{b}_a  ~ \rightarrow  ~ \partial_b \pi^{ab}
\label{eq:reduce2}
\end{equation}
In this regard, a useful identity is
\begin{equation}
D_b ( \pi^b_a  )
  =  \partial_b \pi^b_a 
 -  \frac{1}{2} \left[ \pi^{be} \partial_a \gamma_{be} \right]
+ \frac{1}{2} \pi_a^c  ~ Tr ( \partial_c \ln (\gamma)) 
\label{eq:jose0}
\end{equation}
as shown in Appendix \ref{appf}. 
Under the orthogonal decomposition for
$\pi^{nb}$ and $\gamma_{nb}$, the term in the square brackets will contain a divergence
which can be completely eliminated, apart
from a $(TT)$ contribution (which is zero in our choice of isotropic coordinates) [Eqs.~$(3.10)-(3.11)$ of \cite{adm2}]. 
The last term in Eq.~(\ref{eq:jose0}) involves the logarithmic derivative of the
determinant of the metric.  
It will vanish if the volume (whose
element is proportional to this term) is {\em fixed} within ADM.  Although
this is the case in many applications of ADM, this could be in doubt in
\eg cosmological studies of an expanding universe.  However, it can be
justified if \eg the Taub function $\alpha$ of Eq.~(\ref{eq:taub}) is unit
or a constant.
Moreover, for the last part of Eq.~(\ref{eq:reduce2}), 
the difference between
$\partial _b \pi^b_a $ and  $\partial _b \pi^{ab} $ is also a 
divergence under this orthogonal decomposition [Eq.~$(3.12)$ of \cite{adm2}].
Kimura apparently uses this result in the transition from his Eq.~$(2.3b)$ 
to Eq.~$(3.5b)$ in Ref.~\cite{r7},
\[
-2 D_b (\pi^{b}_a ) ~ \rightarrow ~ -2 \partial_b \pi^{ab}
\]
(modulo the raising/lowering operation of indices with the metric).  However,
 Ohta {\em et al.} does not, and instead computes the explicit Christoffel
symbol for the covariant derivative [Eq.~$(3.5)$ of \cite{r8}].  Yet, for a matter Lagrangian of Eq.~(\ref{eq:LM}) without external fields, both obtain the same solutions using an iterative approximation scheme for the lead term of the metric [Eq.~ $(3.9)$ of \cite{r7} and Eq.~$(3.7)$ of  \cite{r8}],
\begin{equation}
{h^T} \approx  \sum_i \frac{ m_i}{4 \pi r_i } \label{eq:kim1}
\end{equation}
where $r_i = | \vecs{r} - \vecs{z}_i |$ and for the momenta 
[Eq.~$(3.13)$ of \cite{r7},\footnote{There is a small typo in Kimura's paper\cite{r7} concerning
the last term on the far right of his Eq.~ $(3.13)$ which should be $r_a$ not $r_{ab}$.  This is rectified in his paper coauthored with T. Ohta\cite{r8}.  },
and Eq.~$(3.10)$ of \cite{r8}],
\begin{equation}
{\pi^i} \approx \frac{1}{8 \pi }
\sum_i \left\{ p_{ia} \left( \frac{1}{r_i} \right) ~ - ~
\frac{1}{4} p_{ib} \partial_a \partial_b r_i \right\} \label{eq:kim2}
\end{equation}
(though only the solution for $\pi^{ij}$ is relevant here.) 
Here we have only
cited the results (without rederiving them) which 
give us confidence in Eq.~(\ref{eq:jose0}) in its use via the reductions of Ref. \cite{adm2} and the iterative me\-thods for obtaining solutions
by Kimura and Ohta.
Therefore, Eqs.~(\ref{eq:kim1}) and (\ref{eq:kim2}) can serve as initial solutions of the metric using the anzatz (\ref{dila}) for the case $\Psi$ 
(or equivalently small $\kappa$) in a perturbative scheme for small $\kappa$.
Ohta's solutions include a transverse-traceless contribution $g_{ab}^{TT}$ 
[Eq. $(3.11)$ of \cite{r8}].
As noted above, the $(TT)$ contribution 
only affects the metric $\gamma_{ab}$ at second order as shown 
by Ohta {\em et al.} and adds linearly at this order [Eq.~$(4.1)$ of \cite{r8}]. 
We have thus iden\-ti\-fied con\-di\-tions for which the field $\Psi$ need not be injected into the shift-vector equa\-tion
(or, alternatively, conditions for eliminating
most or all of the terms involving $\Pi \partial_a \Psi $), or for which the Christoffel symbols need not be included in the covariant derivative, thus making the  task of solving for the momenta $\pi^{ij}$ tractable in terms of known methods.

\section{Hamiltonian Constraint}

Taking all the simplifications of Sec.~\ref{sec:elim} into account including Eq.~(\ref{eq:curv2}), $K$ in Eq.~(\ref{eq:piiso}),
the isotropic metric of Eq.~(\ref{eq:isotropic}) with $K=0$, and, especially the combination of Eqs.~(\ref{eq:R3p1}) and (\ref{eq:R4})
which allows us to rewrite $R \Psi$, the Hamiltonian constraint
of Eq.~(\ref{eq:H2}) can be rewritten entirely in terms of the metric and $\Psi$ and its time derivatives as
\begin{eqnarray}
& & \frac{\sqrt{\h}}{2} \left( 5 \grad^2 \Psi -  (\grad \Psi )^2 \right)
+ \frac{5}{8 \sqrt{\h} } \grad \Psi \vdot \grad h^T
\nonumber \\
& - & \h^{3/2}  \left\{ \frac{3}{2}
\frac{\partial^2 \Psi}{\partial t^2}
+ \frac{1}{4} \left(
\frac{\partial \Psi}{\partial t}
 \right)^2
+ \frac{1}{2} \T + \Ricci \Psi + \Lambda  \right\} \nonumber \\
& - & \h^{3/2} \left ( \frac{3}{2}  \extra ~ \frac{\partial \Psi}{\partial t} + \frac{1}{2} \extratwo ~ \Psi \right)
-   
{\F}_0
 = 0 \label{eq:Ho}
\end{eqnarray}
where
\begin{eqnarray}
\extra & = & \const ~ = ~ \frac{\partial (\ln \sqrt\gamma ) }{\partial t} - \partial_a N^a
 =  \frac{\partial (\ln \sqrt\gamma ) }{\partial t} \nonumber \\
\extratwo & = & - \frac{2}{\sqrt{\gamma}} \partial_a \left( \sqrt{\gamma} \gamma^{ab} \partial_b N \right) = 0 
\label{eq:extra}
\end{eqnarray}
and where the gradients are now with respect to the Euclidean $3$-metric $\grad_a = \partial_a$ and $\T$ is the right side of Eq.~(\ref{eq:R4}) \ie
$ \T = \frac{\kappa^2}{4} T_{\mu}^{\mu} - 4 \Lambda  $. 
The dot product of the gradients of $\Psi$ and $h^T$ \ie{} $\grad \Psi \vdot \grad h^T$ results from the Christoffel symbols of the covariant Laplacian of Eq.~(\ref{eq:chris}).  
The second derivative of $\Psi$ with respect to time \ie{} $\frac{\partial^2 \Psi}{\partial t^2}$, appears in Eq.~(\ref{eq:Ho}) because the first term in brackets on the right-hand side of Eq.~(\ref{eq:H2})  
for the Hamiltonian constraint, expressed in terms of the extrinsic 
curvature $K_{ab}$ and the Ricci scalar $\Ricci$,
is rewritten in terms of the d'Alembertian of Eq.~(\ref{eq:redbox}) using
the simplification of Eq.~(\ref{eq:curv2}).

Note that we have rewritten the constraint (\ref{eq:H}) in terms of time derivatives of $\Psi$ 
for reasons of convenience: our goal here is to obtain a
self-contained {\em solvable} differential equation for $\Psi$.   Equation (\ref{eq:Ho}) can be recast into
canonical form by employing Eqs.~ (\ref{eq:piiso}) and (\ref{eq:redbox}) to eliminate 
  $\dot{\Psi}$ and $\ddot{\Psi}$ in terms of $\Pi$ and other fields.
Moreover, the term $\extra$ is the coefficient $\const $ of $\dot{\Psi}$
for the d'Alembertian in Eq.~(\ref{eq:redbox}), and $\extratwo$, is the remaining term of the four-dimensional Ricci scalar of Eq.~(\ref{eq:curv})
and is neglected as mentioned in the  simplification of Eq.~(\ref{eq:curv2}). 

Using the separability of Eq.~(\ref{sep2}) for $\Psi$
by which the term $\Pi \partial_a \Psi$ can be eliminated in the shift-vector equation of Sec.~\ref{sec:covector}, we eliminate
$ (\grad \Psi )^2 $ using the same approach as in Eq.~$(29)$ of \cite{mann:gravity1}, \ie:
\begin{equation}
\Psi(t,x,y,z) = F(t) -c \ln ( | \psi (x,y,z) | ) \label{eq:sep}
\end{equation}
With $c=5$ Eq.~(\ref{eq:Ho}) becomes
\begin{eqnarray}
&&  - \frac{25\sqrt{\h}}{2}  \frac{\grad^2 \psi } {\psi }
- \frac{25}{8 \sqrt{\h} } \frac{\grad \psi \vdot \grad h^T }{\psi }
\nonumber \\
& - & \h^{3/2} \left\{ \frac{3}{2} \frac{\partial^2 F(t)}{\partial t^2}
+ \frac{1}{4} \left( \frac{\partial F(t)}{\partial t} \right)^2
+ \frac{3}{2} \extra \frac{\partial F(t)}{\partial t}
\right\}
 \label{eq:H1}  \\
& - &  \frac{\h^{3/2}}{2} \left( \T + 
\Ricci 
~ ( 2 F(t) - 10 \ln (|\psi | ))  
+ 2 \Lambda  \right) - {\F}_0 ~.
= 0
\nonumber
 \end{eqnarray}
If we divide the above by $\h^{3/2}$ and ignore  $\Ricci$, Eq.~(\ref{eq:H1}) divides into the sum of a pure function of $t$ only and a function of the spatial coordinates only
where each term is forcibly a constant (for all time and spatial positions) depending of course on how the matter Lagrangian term depends on spacetime coordinates. 

For further simplicity, let us consider a matter Lagrangian term depending only on the spatial coordinates.  If we let $F(t) = F$ be a con\-stant,
$\mbox{Eq.}$~(\ref{eq:sep}) fits in\-to the pattern of Eq.~(\ref{dila}) and  $\phi$ is time independent 
[and terms like $\ddot{\Psi}$ do not appear in Eq.~(\ref{eq:Ho})].
Also, the term in $\extra $ of
Eq.~(\ref{eq:extra}) from the coefficient of $\dot{\Psi}$ in Eq.~(\ref{eq:redbox}) drops out, as mentioned before.
Note that in the $1+1$ case, $c=4$ and the difference is caused by the first term coupled to $\Psi$ in Eq.~(\ref{eq:H2}), \ie
$ \left( K_{ab} K^{ab} - K^2 - \Ricci \right) $, which does not appear in the $1+1$ case. The term $\grad \psi \vdot \grad h^T$ vanishes
in the far field; it can be eli\-mi\-na\-ted via the transformation
\begin{equation}
\wave ~ = ~ \h^{1/4} \psi \label{eq:rel}
\end{equation}
and Eq.~(\ref{eq:H1}) can be rewritten as
\begin{equation}
-\frac{1}{2} \grad^2 \wave + V \wave + S \wave \ln (| \wave | ) -  E \wave =0
\label{eq:Hf}
\end{equation}
where
\begin{eqnarray}
V & = & \frac{1}{16 \h } \grad^2 h^T - \frac{3}{128 \h^2 } (\grad h^T )^2 - \frac{\h}{50} \T  \nonumber \\
& - &\h  \left( \frac{\ln (\h )}{20}   + \frac{F}{25} \right)  
\Ricci 
- \frac{ h^T \Lambda}{50} - \frac{1}{25 \sqrt{ \h}} \; 
{\F}_0
 ~, \nonumber \\
E & = &  -\frac{ \Lambda }{25}  \quad \mbox{and} \quad S ~ = ~ \frac{\h}{5} \; 
\Ricci
~. \nonumber
\end{eqnarray}
Eq.~(\ref{eq:Hf}) has the functional form of a {\em logarithmic Schr\"{o}dinger equation}.  
As mentioned before,
the term $\extratwo$ of Eq.~(\ref{eq:extra}) resulting from the 4-dimensional scalar curvature 
of Eq.~(\ref{eq:curv}) merely adds to the
the Ricci scalar of the 3-metric. 
The Ricci scalar is
\begin{equation}
\Ricci ~ = ~ \frac{1}{\h^2} \grad^2 h^T - \frac{3}{8 \h^3 } (\grad h^T ) ^2 \label{eq:Ricci3}
\end{equation}
Thus, the potential $V$ in Eq.~(\ref{eq:Hf}) is made of gradients of the metric and the matter Lagrangian term as well as the gravitational constant
(common to the ``eigenenergy'' $E$).

\subsection{Equation for the $3$-metric}

There remains the matter of obtaining $h^T$ for the isotropic coordinates themselves.   The relationship between $p^{ab}$ and $K^{ab}$ in Eq.~(\ref{eq:pK}) with
 Eq.~(\ref{eq:R4}) and Eq.~(\ref{eq:curv2}), the latter resulting from
 Eq.~(\ref{simp}), becomes
\begin{eqnarray}
\Psi^2 R & = &  \Psi^2 \MyBox \Psi  =  \Psi^2 (K^{ab} K_{ab} - K^2 + \Ricci )
\\
& = &
\frac{4}{\gamma} \left( \pi_{ab} \pi^{ab} - \frac{1}{2} \pi^2 \right) 
+ \frac{3}{2} \dot{\Psi}^2 + 2 \Psi K \dot{\Psi} 
+ \Ricci ~ \Psi^2  \nonumber \\
& = &
\frac{4}{\gamma} \left( \pi_{ab} \pi^{ab} - \frac{1}{2} \pi^2 \right)
+ \frac{3}{2} \dot{\Psi}^2 + \Ricci ~ \Psi^2  \nonumber
\end{eqnarray}
since $K=\dot{K}=0$.
Using the same simplifications as before, we obtain
\begin{eqnarray}
 & & 5 \h^2 \left( \frac{\grad^2 \psi }{\psi} - \frac{ (\grad \psi )^2}{\psi^2 } \right)
+ \frac{4}{25}  \frac{(\pi^{ab} \pi_{ab} - \frac{1}{2} \pi^2 )}{ \ln (|\psi| )^2 }
\label{eq:cons}
  \\
 &&\quad + \h \left( \grad^2 h^T +
 \frac{5}{4} \frac{ \grad \psi \vdot \grad h^T }{\psi } \right)
 - \frac{3}{8} (\grad h^T )^2  
~= ~ 0.
 \nonumber
\end{eqnarray}
with $\psi$ is related to $\wave$ via Eq.~(\ref{eq:rel}),  where Eq.~(\ref{eq:Ricci3}) for the Ricci scalar of the 3-metric was used,
and again the gradients in Eq.~(\ref{eq:cons}) are with respect to to 
Euclidean $3$-metric.
Eq.~(\ref{eq:cons}) looks complicated, but once $\pi^{ab}$ is obtained from the shift-vector equation and $\wave$ from the
Hamiltonian, it is a matter of injecting these quantities 
into Eq.~(\ref{eq:cons}) through 
which we can solve for the metric term $h^T$.
In the far field, the coefficient of the log potential term $S \rightarrow 0$ and consequently Eq.~(\ref{eq:Hf})
reduces to the standard linear Schr\"{o}dinger equation and
is completely decoupled from the metric terms $\h$ and $h^T$ that also go to zero, and thus asymptotically,
the equation governing the dilaton
field is completely self-contained with the matter Lagrangian (subject to regularization with respect to the metric term $\h$)
being the case also for the $1+1$ problem \cite{mann:gravity2}.

\section{Discussion}

After deriving the field equations for the scalar-tensor theory represented by Eq.~(\ref{Lag1}),
a $d+1$-dimensional version of the $1+1$ action of Eq.~(\ref{e1}) was proposed and shown to yield the
cor\-rect nonre\-la\-ti\-vis\-tic limit in $d$ di\-men\-sions. We then applied the Arnowitt-Deser-Misner 
method with the gauge settings of Eq.~(\ref{simp}) and the or\-tho\-gonal de\-com\-po\-si\-tions of the
momentum $\pi^{ab}$ and the metric $g_{ab}$  according to Eq.~(\ref{eq:decomp}).  
We found
essentially three coupled PDEs:  the shift-co\-vec\-tor equa\-tion governing the momentum
$\pi^{ab}$ as given by Eq.~(\ref{eq:VN}) and discussed in 
sec.~(\ref{sec:covector}),
the Hamiltonian constraint governing the dilaton field $\Psi$ as given by 
Eq.~(\ref{eq:H}) [or equivalently Eq.~(\ref{eq:H2})],
and a relationship
by which the metric can be obtained from the solutions of 
$\pi^{ab}$ and $\Psi$ namely Eqs.~(\ref{eq:R3p1}) and (\ref{eq:curv}) with Eq.~(\ref{eq:pK}).
 
We found that under the right choice of coordinate and gauge conditions, \ie{} in isotropic coordinates and when $g^{TT}_{ab}= h^{TT}_{ab}=0$,
the PDE go\-ver\-ning the dilaton field in Eq.~(\ref{eq:H2}) is
a lo\-ga\-rith\-mic Schr\"{o}dinger equa\-tion, the nonlinear logarithmic term being directly proportional
 to the Ricci scalar of the $3$-metric which becomes zero in the far-field limit (Minkowski flat-space) as given by (\ref{eq:Hf}).
In the latter regime, the PDE becomes a linear Schr\"{o}dinger equation in the far field. 
Equation
(\ref{eq:cons}) then allows one to 
solve for the metric in terms of $\psi$.


Thus, we find that our proposed $3+1$ scalar-tensor theory holds similar properties to those
of the $1+1$ formulation, very much what would be expected from sound dimensional scaling.
The main difference is that unlike the $1+1$ case where the reduced Hamiltonian is
expressed as a form of spatial integral of the second derivative of the
scalar field, the $3+1$ formulation uses the second derivative of the metric function
$h^T$, just as in standard GRT.  This can be understood from the fact that
in $1+1$ di\-men\-sions, the dilaton field contributes most of the physics (GRT in $1+1$ dimensions yields nothing). However,
in $3+1$ di\-men\-sions, the dilaton field represents a small departure from standard GRT.

This outcome is interesting as the logarithmic Schr\"{o}dinger equation
finds applications in quantum mechanics, the theory of superfluidity and Bose-Einstein condensates \cite{bose},
and even nuclear physics \cite{nuclear}.
As in the previous $1+1$ case, 
since we already know the Lagrangian density whose Euler-Lagrange equations are the
(linear) Schr\"{o}dinger wave equation, we can obtain a Hamiltonian density and quantize the system.
This procedure, often called second quantization,  allows {\em transitions} between states, with
the dilaton itself acting as the agent of transition.  In effect, this yields a theory
of quantum gravity which is normalizable because {\em only} the dilaton field is quantized.
(Note that our treatment of the one-graviton exchange in Appendix \ref{appa} is still relevant because
the effective potentials $V_d$ are obtained in the lowest order and have a well-known correspondence with
classical mechanics in the limit $\hbar \rightarrow 0$.)

From here, many di\-rec\-tions are possible. An ob\-vious next step is to ex\-pli\-ci\-tly solve Eq.~(\ref{eq:ppi})
for the momentum $\pi_{ab}$ according to sec. \ref{sec:covector}, Eq.~(\ref{eq:Hf})
for the wave function $\wave$, and Eq.~(\ref{eq:cons}) for the metric term $h^T$  to obtain
the Hamiltonian of Eq.~(\ref{eq:energy}) using the explicit terms of ${\F}_{\mu}$ from Eq.~(\ref{eq:F}) \ie{} 
the various components of the matter Lagrangian
in Eq.~(\ref{eq:LM}).  
Analytical solutions are desirable, and we anticipate that the
generalized Lambert $W$ function will be useful
as it was for the $1+1$ lineal gravity problem\cite{scott:aaecc,scott:sigsam1,scott:sigsam2}.
Departures from the gauge conditions of Eq.~(\ref{simp}) and the addition 
of a nonzero
transverse-traceless component for the metric (to model gravitational radiation,perhaps)
can be explored by iterative schemes like those mentioned in Sec. \ref{sec:covector}.

It has been hypothesized that superfluid vacuum theory (SVT) might be responsible for the mass mechanism
(in contradistinction to the Higgs boson or perhaps working in tandem with it). Some versions of SVT favor a
logarithmic Schr\"{o}dinger equation \cite{svt}.  Given the apparent resemblance between the Higgs
boson and the dilaton mentioned earlier in the Introduction, the formulation herein could be fruitful in
investigating this direction.  However, we wish to emphasize that regardless of whether the
dilaton and Higgs boson are related to each other or not, the results of the present work and their
implications concerning quantum gravity stand on their own.

The very fact of the dilaton field being
governed by an energy-balancing quantum mechanical wave equation suggests that the wave function
itself might also be a ``geometrical'' quantity, apart from its usual 
interpretation; this would be
very much as spin is treated in GRT, but such a notion needs further investigation.
Granted, we have found a particular class of solutions to the scalar-tensor gravitational theory
proposed in Eq.~(\ref{Lag1});  nonetheless, the class 
found here shares similar essential properties
with its simpler $1+1$ counterpart of Eq.~(\ref{e1}).

\begin{acknowledgments}
First, we would like to thank the departed Professor Tadayuki Ohta
who laid the foundations of this work in $1+1$ di\-men\-sions.  We would
also like to thank the following individuals:
Jos\'{e} Miguel Figueroa-O'Farrill of the University of Edinburgh
and Sadri Hassani formerly of the University of Illinois at Urbana-Champaign,
Manuel Tiglio for helpful discussions
on the Taub function, Andrew Zador, president of Fovix Corp. Canada, Kent
Nickerson, retired Principal Scientist of RF, Blackberry,
Johannes Grotendorst of the Forschungszentrum J\"{u}lich,
Madhusudan Therani, Principal of EngKraft LLC, and Sudarshan Therani Nadathur of Antaryami Systems Consulting Private Limited (Chennai , India).
Finally, we would also like to thank Professor Yuncai Wang of
TYUT for supporting this work.
Professor Wang is supported by the National Natural Science Foundation of China
under Grants No. 61405138, No. 61475111, and No. 61227016, by the Program
for the Outstanding Innovative Teams of Higher Learning
Institutions of Shanxi.
T.C. Scott is supported in China by the project GDW201400042 for the ``high end foreign experts project''. 
R.B. Mann was supported in part by the Natural Sciences and Engineering Research Council of Canada.
\end{acknowledgments}

\appendix

\section{Correspondence with Newtonian gravity in $d+1$ di\-men\-sions}
\label{appa}
In this section we illustrate how a Newtonian limit ge\-ne\-ri\-cally arises in
$d+1$ di\-men\-sions for the sca\-lar-ten\-sor theory of Eq.~(\ref{Lag1}).
We shall compute the New\-tonian limit(s) by considering the one gra\-vi\-ton
exchange potential (keeping in mind that there are no propagating gra\-vi\-tons
in two or three space\-time di\-men\-sions).  Gra\-vi\-tons can only propagate in $3+1$
di\-men\-sions (so the treatment for $n=d+1 < 4$ is formal).

Specifically, we calculate  the $T$-matrix element of the one gra\-vi\-ton exchange
diagram in $d+1$ Einstein gravity in the
framework of the conventional quantum field theory, and we determine
the classical potential as the Fourier transformation of the $T$-matrix
element in the limit of $h \rightarrow 0$.
Here the word ``conventional'' means that we do not touch on the
Faddev-Popov (FP) ghosts etc \ldots As far as we treat the lowest order and static
contributions, this causes no problem.

We begin by extending the theory in (\ref{e1}) to $d+1=n$ di\-men\-sions and
coupling $N$ scalar fields.  We introduce the neutral scalar fields with mass $m_{a}$ instead of
the point particles. This yields the Lagrangian density:
in $1+1$ dimensions the $\tilde{g}^{\mu\nu}\equiv \sqrt{-g}g^{\mu\nu}$ is not
an appropriate variable for developing the quantum field theory,
because its components are not independent due to the identity
$\mbox{det}(\tilde{g}^{\mu\nu})=-1$.
We define the gra\-vi\-ton field $h_{\mu\nu}$
and the dilaton field $\psi$ from Eq.~(\ref{dila}) via
\begin{equation}
g_{\mu\nu}=\eta_{\mu\nu}+\kappa h_{\mu\nu} \label{gravi}
\end{equation}
where $\eta_{\mu \nu}$ is the metric for flat (Minkowski) space.
The reason why we defined the dilaton field not by $\Psi=\kappa\psi$ but
by (\ref{dila}) is to introduce the correct kinematical part of the
gra\-vi\-ton field and also formally ensure the Einstein-Hilbert action
as $\kappa \rightarrow 0$ for $\Psi$.

Though this separation in Eq.~(\ref{gravi}) is well known in the weak-field
approximation or  ``linearized gravity'' \cite{radiation}, it
is exact and can be done without any loss of
generality. However the counterpart in terms of upper indices is not
exactly separable.  To first order in $\kappa$, it is gi\-ven by
\begin{equation}
g^{\mu\nu} \approx \eta^{\mu\nu}+\kappa h^{\mu\nu} \label{eq:upper}
\end{equation}
where
$
h^{\mu \nu } = \eta^{\alpha \mu } \eta^{\beta \nu} h_{\alpha \beta}
$
and we treat $h^{\mu\nu}$ as the gra\-vi\-ton field.
In general, it is not possible to separate the metric into a static background and a perturbation in the form of radiation.
However, if such a separation can be made
the perturbation is transverse (perpendicular to the direction of motion)
and traceless relative to the background.  Thus the
physical gra\-vi\-ton is the transverse-traceless part of $h^{\mu\nu}$,
we refer here to all components as gra\-vi\-ton.  Here, the Lagrangian density of Eq.~(\ref{Lag1}) becomes
\begin{eqnarray}
L_{0}&=&-\frac{1}{2}\left\{\partial_{\lambda}h^{\mu\nu}\partial^{\lambda}
h_{\mu\nu}-\partial^{\lambda}h^{\mu}_{\;\;\mu}
\partial_{\lambda}h^{\nu}_{\;\;\nu}
-2\partial_{\mu}h^{\lambda}_{\;\;\nu}\partial_{\lambda}h^{\mu\nu} \right.
\nonumber \\
&& \left. +2\partial_{\mu}h^{\mu\nu}\partial_{\nu}h^{\lambda}_{\;\;\lambda}\right\}
 +2\partial^{\mu}\left(\partial^{\nu}h_{\mu\nu}
-\partial_{\mu}h^{\nu}_{\;\nu}\right)\psi
+\partial^{\mu}\psi\partial_{\mu}\psi
\nonumber \\
&& -\frac{1}{2}\sum_{a}\left(\pha^{,\;\mu}_{a}\pha_{a,\;\mu}
+m^{2}_{a}\pha^{2}_{a}\right)\;\;.
\end{eqnarray}
where the cosmological constant $\Lambda$ has been set to zero.
In the spirit of Feynman's
path integral approach \cite{feynman}, [Sec.~$(1.2)$ of \cite{hamber}], only products of at most two $h$ terms and first order
derivatives of each term in $h$ are desired in the Lagrangian.  Thus,
we rewrite this latter in the form
\begin{eqnarray}\label{Lag2}
L_{0}&=&-\frac{1}{2}\left\{\partial_{\lambda}h^{\mu\nu}\partial^{\lambda}
h_{\mu\nu}-\partial^{\lambda}h^{\mu}_{\;\;\mu}
\partial_{\lambda}h^{\nu}_{\;\;\nu}
-2\partial_{\mu}h^{\mu\nu}\partial^{\lambda}h_{\lambda\nu} \right.
\nonumber \\
&& \left. +2\partial_{\mu}h^{\mu\nu}\partial_{\nu}h^{\lambda}_{\;\;\lambda}\right\}
-2\left(\partial^{\nu}h_{\mu\nu}
-\partial_{\mu}h^{\nu}_{\;\nu}\right)\partial^{\mu}\psi
+\partial^{\mu}\psi\partial_{\mu}\psi
\nonumber \\
&&-\frac{1}{2}\sum_{a}\left(\pha^{,\;\mu}_{a}\pha_{a,\;\mu}
+m^{2}_{a}\pha^{2}_{a}\right)\;\;.
\end{eqnarray}
where we dropped divergences (which contribute nothing to the Euler-Lagrange equa\-tions)
for the free Lagrangian density following from (\ref{Lag1}).
The characteristic of this Lagrangian density is the mixing term
between gra\-vi\-ton and dilaton. However, we can eliminate this by
redefining the dilaton field as
\begin{equation}
\tilde{\psi}\equiv \psi+h^{\mu}_{\;\mu}-\frac{\partial^{\mu}\partial^{\nu}}
{\MyBox}h_{\mu\nu}
\end{equation}
which allows us to express (\ref{Lag2}) as
\begin{eqnarray}
\label{Lag3}
L_{0} &=& -\frac{1}{2}\left\{\partial_{\lambda}h^{\mu\nu}\partial^{\lambda}
h_{\mu\nu}-\partial^{\lambda}h^{\mu}_{\;\;\mu}
\partial_{\lambda}h^{\nu}_{\;\;\nu}
-2\partial_{\mu}h^{\mu\nu}\partial^{\lambda}h_{\lambda\nu} \right.
\nonumber \\
&& \left. +2\partial_{\mu}h^{\mu\nu}\partial_{\nu}h^{\lambda}_{\;\;\lambda}\right\}
 -\left(\partial^{\nu}h_{\mu\nu}-\partial_{\mu}h^{\nu}_{\;\nu}\right)
\left(\partial^{\lambda}h^{\mu}_{\;\lambda}
-\partial^{\mu}h^{\lambda}_{\;\lambda}\right)
\nonumber \\
&&+\partial_{\mu}\tilde{\psi}\partial^{\mu}\tilde{\psi}
-\frac{1}{2}\sum_{a}\left(\pha^{,\;\mu}_{a}\pha_{a,\;\mu}
+m^{2}_{a}\pha^{2}_{a}\right)\;\;.
\end{eqnarray}
The field $\tilde{\psi}$ decouples from the Lagrangian and we shall not
consider it further. The free Lagrangian density of the gra\-vi\-ton is obtained by simplifying
the first two terms above to obtain
\begin{eqnarray}
L_{0g}&=&-\frac{1}{2}\left\{\partial_{\lambda}h^{\mu\nu}\partial^{\lambda}
h_{\mu\nu}+\partial^{\lambda}h^{\mu}_{\;\;\mu}
\partial_{\lambda}h^{\nu}_{\;\;\nu}
-2\partial_{\mu}h^{\mu\nu}\partial_{\nu}h^{\lambda}_{\;\;\lambda}\right\}
\nonumber \\
&& \quad +\partial^{\nu}h_{\mu\nu}B^{\mu}+\frac{1}{4}B_{\mu}B^{\mu}
\end{eqnarray}
where we added gauge fixing terms in the form of a Lagrange multiplier
field $B_{\mu}$ in the Feynman gauge (\eg see Feynman rules [Sec.~$(1.4)$ of \cite{hamber}]). We
solve for $B_{\mu}$ directly from $L_{0g}$  in terms of the field equa\-tion
\[
\partial^{\nu}h_{\nu\mu} +\frac{1}{2}B_{\mu}=0 ~.
\]
Thus $B_{\mu} = -2\partial^{\nu}h_{\nu\mu} $ and $B^{\mu} \approx -2\partial_{\nu}h^{\nu\mu} $
with respect to the background metric $\eta_{\mu \nu }$.
Eliminating $B_{\mu}$ and $B^{\mu}$ from its field equa\-tion leaves us with the
Lagrangian
\begin{eqnarray}\label{Lag4}
\tilde{L}_{0g}&=&-\frac{1}{2}\left\{\partial_{\lambda}h^{\mu\nu}\partial^{\lambda}h_{\mu\nu}+\partial^{\lambda}h^{\mu}_{\;\;\mu}
\partial_{\lambda}h^{\nu}_{\;\;\nu}
-2\partial_{\mu}h^{\mu\nu}\partial_{\nu}h^{\lambda}_{\;\;\lambda}\right\}
\nonumber\\
&& \quad -\partial^{\nu}h_{\mu\nu}\partial_{\lambda}h^{\mu\lambda}\;\;.
\end{eqnarray}
whose canonical quantization we shall now undertake.

Temporarily setting the scalar fields to zero, we obtain
\begin{eqnarray}\label{feq1a}
&&\MyBox h_{\mu\nu}+\eta_{\mu\nu}\MyBox h^{\lambda}_{\;\;\lambda}
-\eta_{\mu\nu}\partial_{\lambda}\partial_{\rho}h^{\lambda\rho}
-\partial_{\mu}\partial_{\nu}h^{\lambda}_{\;\;\lambda}
\nonumber \\
&& \quad +\partial_{\mu}\partial^{\lambda}h_{\nu\lambda}
+\partial_{\nu}\partial^{\lambda}h_{\mu\lambda}=0\;\;.
\end{eqnarray}
for the gra\-vi\-ton field equa\-tion. Its trace is
\begin{equation}
\MyBox h^{\lambda}_{\;\;\lambda}=
\frac{n-2}{n}\partial_{\lambda}\partial_{\rho} h^{\lambda\rho} ~. \label{feq1as}
\end{equation}
In $1+1$ di\-men\-sions, $n=2$ and $\MyBox h^{\lambda}_{\;\;\lambda}=0$.
In general, Eq.~(\ref{feq1as}) implies that (\ref{feq1a}) becomes
\begin{equation}\label{feq2a}
\MyBox h_{\mu\nu}
-\frac{2}{n}\eta_{\mu\nu}\partial_{\lambda}\partial_{\rho}h^{\lambda\rho}
-\partial_{\mu}\partial_{\nu}h^{\lambda}_{\;\;\lambda}
+\partial_{\mu}\partial^{\lambda}h_{\nu\lambda}
+\partial_{\nu}\partial^{\lambda}h_{\mu\lambda}=0\;\;.
\end{equation}
Taking  the $\partial^{\nu}$ derivative of (\ref{feq2a}) leads to
\begin{equation}\label{feq3a}
\MyBox\partial^{\nu}h_{\mu\nu}=0\;\; .
\end{equation}
This, along with the d'Alembertians of (\ref{feq2a})
and (\ref{feq3a}) respectively lead to
\begin{equation}\label{feqAA}
{\MyBox}^{2} h_{\mu \nu} -\partial_{\mu} \partial_{\nu} \MyBox h^{\lambda}_{\;\;\lambda}
=0\;\; \quad \mbox{and} \quad   {\MyBox}^{2}    h^{\lambda}_{\;\;\lambda}=0\;\;
\end{equation}
which finally implies
\begin{equation}\label{feq4a}
{\MyBox}^{3} h_{\mu\nu}=0
\end{equation}
a relation characteristic of $n\geq 3$ Lagrangian.  For $n=2$, we simply get
${\MyBox} h_{\mu\nu}=0$.
The conjugate momentum is:
\begin{eqnarray}\label{mom}
\pi^{\mu\nu}&=&\frac{\partial L_{0}}
{\partial(\partial_{0}\gamma_{\mu\nu})}
\nonumber \\
&=&\partial_{0} h^{\mu\nu}+\eta^{\mu\nu}
\left(\partial_{0} h^{\lambda}_{\;\;\lambda}
+\partial_{\lambda} h^{\lambda}_{\;\;0}\right)
+\eta^{\mu 0}\left(\frac{1}{2}\partial^{\nu}h^{\lambda}_{\;\;\lambda} \right.
\nonumber \\
&&- \left. \partial_{\lambda} h^{\nu\lambda}\right)
+\eta^{\nu 0}\left( \frac{1}{2}\partial^{\mu}h^{\lambda}_{\;\;\lambda}
-\partial_{\lambda} h^{\mu\lambda} \right)\;\;.
\end{eqnarray}
which implies
\begin{eqnarray}\label{comp2}
\partial_{0}h_{00}&=&\frac{1}{2}\pi^{00}+\frac{1}{2}\partial_{i}h_{0i}
\nonumber \\
\partial_{0}h_{0i}&=&-\frac{1}{2}\pi^{0i}+\frac{1}{4}\partial_{i}h_{00}
-\frac{1}{4}\partial_{i}h_{jj}+\frac{1}{2}\partial_{j}h_{ij}
\\
\partial_{0}h_{ij}&=&\pi^{ij}-\frac{\delta_{ij}}{n}\pi^{kk}
+\frac{\delta_{ij}}{n}\partial_{k}h_{0k}\;\;.
\nonumber
\end{eqnarray}
The equal-time commutation relations are
\begin{eqnarray}
\left[h_{\mu\nu}(x), \;\pi^{\lambda\rho}(y)\right]_{eq}
&=&\frac{i}{2}\left(\delta^{\lambda}_{\mu}\delta^{\rho}_{\nu}
+\delta^{\rho}_{\mu}\delta^{\lambda}_{\nu}\right)\delta^{(n-1)}(x-y)
\nonumber \\
\\
\left[h_{\mu\nu}(x), \;h_{\lambda\rho}(y)\right]_{eq}
&=&\left[\pi^{\mu\nu}(x), \;\pi^{\lambda\rho}(y)\right]_{eq}=0\;\;.
\nonumber
\end{eqnarray}
implying that the commutators between $h_{\mu\nu}$ and
$\partial_{0}h_{\lambda\rho}$ become
\begin{eqnarray}\label{commu1}
&&\left[h_{\mu\nu},\;\partial_{0}h_{\kappa\sigma}\right]_{eq}
\nonumber \\
&&=\frac{i}{2}\left\{\frac{1}{2}\left(\eta_{\mu\kappa}\eta_{\nu\sigma}
+\eta_{\mu\sigma}\eta_{\nu\kappa}\right)
+\frac{1}{2}\left(\bar{\eta}_{\mu\kappa}\bar{\eta}_{\nu\sigma}
+\bar{\eta}_{\mu\sigma}\bar{\eta}_{\nu\kappa}\right) \right.
\nonumber \\
&& \qquad \left. -\frac{2}{n}\bar{\eta}_{\mu\nu}\bar{\eta}_{\kappa\sigma}\right\}
\delta^{(n-1)}(x-y)
\end{eqnarray}
where
\begin{equation}
\bar{\eta}_{\mu\nu}\equiv \eta_{\mu\nu}+\eta_{\mu 0}\eta_{\nu 0}\;\;.
\end{equation}
The proof of this relation is gi\-ven in the Appendix of ref.~\cite{mann:gravity1}.
The solution to (\ref{feq4a}) is
\begin{eqnarray}\label{h}
h_{\mu\nu}(x)&=&-\int d^{n-1}z\;D^{(n)}(x-z)\bar{\partial}^{z}_{0}h_{\mu\nu}(z)
\nonumber \\
&&-\int d^{n-1}z\;\tilde{D}^{(n)}(x-z)\bar{\partial}^{z}_{0}\MyBox h_{\mu\nu}(z)
\nonumber \\
&&-\int d^{n-1}z\;\tilde{\tilde{D}}^{(n)}(x-z)\bar{\partial}^{z}_{0}{\MyBox}^{2}
h_{\mu\nu}(z)
\;\;.
\end{eqnarray}
where  $D^{(n)}$, $\tilde{D}^{(n)}$ and $\tilde{\tilde{D}}^{(n)}$ are
defined via
\begin{eqnarray}
D^{(n)}(x)
&=&-\frac{i}{(2\pi)^{n-1}}\int d^{n}k\;\epsilon(k_{0})\delta(k^{2})e^{ikx}
\nonumber \\
\tilde{D}^{(n)}(x)
&=&-\frac{i}{(2\pi)^{n-1}}\int d^{n}k\;\epsilon(k_{0})\delta^{\prime}
(k^{2})e^{ikx}
\nonumber \\
\tilde{\tilde{D}}^{(n)}(x)
&=&-\frac{i}{(2\pi)^{n-1}}\int d^{n}k\;\epsilon(k_{0})\delta^{\prime\prime}
(k^{2})e^{ikx}
\nonumber
\end{eqnarray}
as elaborated in appendix \ref{appb}.
We next need to express all of $h_{\mu\nu}, \partial_{0}h_{\mu\nu},
\MyBox \partial_{0}h_{\mu\nu}, {\MyBox}^{2}h_{\mu\nu}$ and
${\MyBox}^{2}\partial_{0}h_{\mu\nu}$ in terms of the canonical variables
and calculate commutators at equal-time.  This rather lengthy and
complicated calculation is gi\-ven in the Appendix of ref.~\cite{mann:gravity1}.

From (\ref{h}) and the equal-time commutators,
the commutator among the components of $h_{\mu\nu}$ at two
arbitrary space-time points can be calculated
\begin{eqnarray}\label{commu2}
\lefteqn{\left[h_{\mu\nu}(x),\;h_{\lambda\rho}(y)\right]}
\nonumber \\
&&=\frac{i}{2}\left(\eta_{\mu\lambda}\eta_{\nu\rho}
+\eta_{\mu\rho}\eta_{\nu\lambda}
-\frac{2}{n}\eta_{\mu\nu}\eta_{\lambda\rho}\right)D^{(n)}(x-y)
\nonumber \\
&&+\frac{i}{4}\left\{-\eta_{\mu\lambda}\partial_{\nu}\partial_{\rho}
-\eta_{\mu\rho}\partial_{\nu}\partial_{\lambda}
-\eta_{\nu\lambda}\partial_{\mu}\partial_{\rho}
-\eta_{\nu\rho}\partial_{\mu}\partial_{\lambda} \right.
\nonumber \\
&&+ \left. \frac{4}{n}\left(\eta_{\mu\nu}\partial_{\lambda}\partial_{\rho}
+\eta_{\lambda\rho}\partial_{\mu}\partial_{\nu}\right)\right\}
\tilde{D}^{(n)}(x-y)
\nonumber \\
&&+\frac{i}{2}\left(1-\frac{2}{n}\right)\partial_{\mu}\partial_{\nu}
\partial_{\lambda}\partial_{\rho}\tilde{\tilde{D}}^{(n)}(x-y)\;\;.
\end{eqnarray}
This expression is valid even when $n=2$.
(The proof is also gi\-ven in the Appendix of \cite{mann:gravity1}).
The gra\-vi\-ton propagator is
\begin{equation}
\langle 0|T(h_{\mu\nu}(x)h_{\lambda\rho}(y))|0\rangle
=-\frac{i}{2(2\pi)^{n}}
\int d^{n}k\;e^{ik(x-y)}\frac{X_{\mu\nu,\lambda\rho}}{k^{2}-i\epsilon}
\label{eq:propagator}
\end{equation}
where
\begin{eqnarray}\label{propag}
X_{\mu\nu,\lambda\rho}&=&\eta_{\mu\lambda}\eta_{\nu\rho}
+\eta_{\mu\rho}\eta_{\nu\lambda}-\frac{2}{n}\eta_{\mu\nu}\eta_{\lambda\rho}
\nonumber \\
&&+\frac{1}{2k^{2}}\left\{-\eta_{\mu\lambda}k_{\nu}k_{\rho}
-\eta_{\mu\rho}k_{\nu}k_{\lambda}-\eta_{\nu\lambda}k_{\mu}k_{\rho}
-\eta_{\nu\rho}k_{\mu}k_{\lambda} \right.
\nonumber \\
&&\quad + \left. \frac{4}{n}\left(\eta_{\mu\nu}k_{\lambda}k_{\rho}
+\eta_{\lambda\rho}k_{\mu}k_{\nu}\right)\right\}
\nonumber \\
&&+\left(1-\frac{2}{n}\right)\frac{k_{\mu}k_{\nu}k_{\lambda}k_{\rho}}
{(k^{2})^{2}}\;\;.
\end{eqnarray}
We turn now to the scalar fields.
Since $\sqrt{-g}=1+\frac{\kappa}{2} h^{\mu}_{\mu}+{\cal O}(\kappa^{2})$.
the interaction Lagrangian in the lowest order is
\begin{equation}
L_{int}=-\frac{1}{2}\left\{\frac{1}{2}\eta^{\mu\nu}
\left(\pha^{,\alpha}\pha_{,\alpha}+m^{2}\pha^{2}\right)
-\pha^{,\mu}\pha^{,\nu}\right\}h_{\mu\nu}\;\;.
\end{equation}
Using the propagator (\ref{eq:propagator}), the $S$-matrix element of the one
gra\-vi\-ton exchange between two scalar particles is calculated as
\begin{widetext}
\begin{eqnarray}
S_n &=& \frac{4\pi iG_{n}}{(2\pi)^{n-2}}
\left(p^{0}_{1}p^{0}_{2}q^{0}_{1}q^{0}_{2}\right)^{-1/2}
\left[p^{\mu}_{1}q^{\nu}_{1}-\frac{1}{2}\eta^{\mu\nu}
(p_{1}\cdot q_{1}+m^{2}_{1})\right]
\left[p^{\lambda}_{2}q^{\rho}_{2}-\frac{1}{2}\eta^{\lambda\rho}
(p_{2}\cdot q_{2}+m^{2}_{2})\right]
\times \frac{X_{\mu\nu,\lambda\rho}}{k^{2}}
\delta^{(n)}(p_{1}+p_{2}-q_{1}-q_{2})
\nonumber \\
S_2&=&4\pi iG \left(p^{0}_{1}p^{0}_{2}q^{0}_{1}q^{0}_{2}\right)^{-1/2}
\frac{1}{k^{2}}\left[(p_{1}\cdot p_{2})(q_{1}\cdot q_{2})
+(p_{1}\cdot q_{2})(q_{1}\cdot p_{1})
-(p_{1}\cdot q_{1})(p_{2}\cdot q_{2})\right]
\times\delta^{(2)}(p_{1}+p_{2}-q_{1}-q_{2})\;\;. \label{eq:Sd}
\end{eqnarray}
\end{widetext}
where $p^{\mu}_{a}$, $q^{\mu}_{a}$ and $k^{\mu}$ are the four-
momenta of the initial particles, the final particles and the
transferred gra\-vi\-ton, respectively. This result is valid for $n=2$ also.
The $T$-matrix element is defined by the formula
\begin{equation}\label{T1}
S_n =-i(2\pi)T_n \delta^{(n)}(p_{1}+p_{2}-q_{1}-q_{2}) ~,
\end{equation}
and the potential is calculated via its Fourier transformation as
\begin{equation}\label{pot1}
V_n =\int d^{n-1}k\;e^{-i{\bf k\cdot r}}T_n ({\bf k}) ~.
\end{equation}
In the lowest order, the static approximation $T$-matrix element
is
\begin{equation}
T_{n}=-4\left(1-\frac{1}{n}\right)\frac{G_{n}}{(2\pi)^{n-2}}
\frac{m_{1}m_{2}}{{\bf k}^{2}}\;\;.
\end{equation}
whose associated potential is $V = \int d^n k e^{-ikx}T(k)$
in $n$ di\-men\-sions.
\ \\
\ \\
\noindent
{\bf Remark:} $k_{\mu}$ in the propagator yields the term as
$(k\cdot p_{1})=-k^{0}p^{0}_{1}+{\bf k\cdot p_{1}}$. This term
does not contribute to the static potential, because $k^{0}$ is
momentum-dependent. The reason is as follows.

Let's consider the energy conservation at the vertices of the one
-gra\-vi\-ton exchange diagram, which leads to
\[
k^{0}=p^{0}_{1}-q^{0}_{1} \quad \mbox{and} \quad
k^{0}=q^{0}_{2}-p^{0}_{2}\;\;.
\]
The choice between these two $k^{0}$ leads to the different result
of momentum sector of the potential. This was pointed out by Y. Nambu
in 1950 \cite{nambu1,nambu2,nambu3}. T. Ohta investigated this problem many years ago and found the
consistent choice is to take an average of two ex\-pres\-sions
\[
k^{0}=\frac{1}{2}(p^{0}_{1}-q^{0}_{1}+q^{0}_{2}-p^{0}_{2})\;\;.
\]
More generally a parameter can be introduced in the above
expression, which was proved to be identical with a gauge parameter.
At any rate $k_{0}$ does not contribute to the static potential.

The $T$-matrix elements for $n=2, 3$ and $4$ are
\begin{eqnarray}
T_{2}&=&-\frac{2G_{2}m_{1}m_{2}}{{\bf k^{2}}}
\\
T_{3}&=&-\frac{8}{3}\cdot\frac{G_{3}}{(2\pi)}\cdot
\frac{m_{1}m_{2}}{{\bf k^{2}}}
\\
T_{4}&=&-\frac{3G_{4}}{(2\pi)^{2}}\cdot\frac{m_{1}m_{2}}{{\bf k^{2}}}\;\;.
\end{eqnarray}
and the corresponding potentials are
\begin{eqnarray}
V_{2}&=&2\pi G_{2}m_{1}m_{2}\;r \label{pot2}
\\
V_{3}&=&2\left(\frac{4}{3}G_{3}\right)m_{1}m_{2}\mbox{log}\;r \label{pot3}
\\
V_{4}&=&-\frac{3}{2}\cdot\frac{G_{4}m_{1}m_{2}}{r} \label{pot4}
\end{eqnarray}
By identifying the gravitational constants as
\begin{equation}
G_{N,2}\equiv G_{2} \qquad G_{N,3}\equiv \frac{4}{3}G_{3}
\qquad G_{N,4}\equiv \frac{3}{2}G_{4}
\end{equation}
we get the correct Newtonian potentials in each dimension. The result for
$V_4$ tells this method has been established and led to the exact potential to the
post-post-Newtonian order.

The results above are in strong contrast with $d+1$-dimensional
GRT, whose free Lagrangian density is
\begin{equation}\label{Lag1o}
{\cal L}=\frac{2}{\kappa^{2}}\sqrt{-g}R-\frac{1}{2}\sum_{a}\sqrt{-g}
(g^{\mu\nu}\pha_{a,\mu}\pha_{a,\nu}+m^{2}_{a}\pha^{2}_{a})
\end{equation}
from which the free Lagrangian of the gra\-vi\-ton is
\begin{eqnarray}
L_{0g}&=&-\frac{1}{2}\left\{\partial_{\lambda}h^{\mu\nu}\partial^{\lambda}
h_{\mu\nu}-\partial^{\lambda}h^{\mu}_{\;\;\mu}
\partial_{\lambda}h^{\nu}_{\;\;\nu}
-2\partial_{\mu}h^{\mu\nu}\partial^{\lambda}h_{\lambda\nu} \right.
\nonumber \\
&&+ \left. 2\partial_{\mu}h^{\mu\nu}\partial_{\nu}h^{\lambda}_{\;\;\lambda}\right\}
\label{sans} \\
&&+\left(\partial^{\nu}h_{\nu\mu}-\frac{1}{2}\partial_{\mu}h^{\lambda}
_{\;\;\lambda}\right)B^{\mu}+\frac{1}{4}B_{\mu}B^{\mu}
\nonumber
\end{eqnarray}
where gauge fixing terms have been added.
A computation analogous to
the one above gives the following. There
are ways of eliminating the gauge fields
\begin{enumerate}
\item Define a new field $C_{\mu}$ by
\[
C_{\mu}\equiv B_{\mu}+2\left(\partial^{\nu}h_{\nu\mu}
-\frac{1}{2}\partial_{\mu}h^{\nu}_{\;\;\nu}\right)
\]
The gauge fixing term of (\ref{sans}) becomes
\begin{eqnarray}
\lefteqn{\left(\partial^{\nu}h_{\nu\mu}-\frac{1}{2}\partial_{\mu}h^{\nu}
_{\;\;\nu}\right)B^{\mu}+\frac{1}{4}B_{\mu}B^{\mu}}
\nonumber \\
&=&-\left(\partial^{\nu}h_{\nu\mu}-\frac{1}{2}\partial_{\mu}h^{\nu}
_{\;\;\nu}\right)\left(\partial^{\lambda}h_{\lambda}^{\mu}
-\frac{1}{2}\partial^{\mu}h^{\lambda}_{\;\;\lambda}\right)
+\frac{1}{4}C_{\mu}C^{\mu}
\nonumber
\end{eqnarray}
The $C_{\mu}$ field is completely separated form the gra\-vi\-ton's world
and has no contribution to physics.
\item Eliminate $B_{\mu}$ directly in terms of the field equa\-tion
\[
\partial^{\nu}h_{\nu\mu}-\frac{1}{2}\partial_{\mu}h^{\nu}_{\;\;\nu}
+\frac{1}{2}B_{\mu}=0
\]
\end{enumerate}
Either way, the gauge fixing term becomes
\begin{eqnarray}
\lefteqn{\left(\partial^{\nu}h_{\nu\mu}-\frac{1}{2}\partial_{\mu}h^{\nu}
_{\;\;\nu}\right)B^{\mu}+\frac{1}{4}B_{\mu}B^{\mu}}
\nonumber \\
&=&-\left(\partial^{\nu}h_{\nu\mu}-\frac{1}{2}\partial_{\mu}h^{\nu}
_{\;\;\nu}\right)\left(\partial^{\lambda}h_{\lambda}^{\mu}
-\frac{1}{2}\partial^{\mu}h^{\lambda}_{\;\;\lambda}\right)
\nonumber
\end{eqnarray}
and the Lagrangian density becomes
\begin{eqnarray}\label{eq2}
L_{0g}
&=&-\frac{1}{2}\left\{\partial_{\lambda}h^{\mu\nu}\partial^{\lambda}
h_{\mu\nu}-\partial^{\lambda}h^{\mu}_{\;\;\mu}
\partial_{\lambda}h^{\nu}_{\;\;\nu}
-2\partial_{\mu}h^{\mu\nu}\partial^{\lambda}h_{\lambda\nu} \right.
\nonumber \\
&&+2 \left. \partial_{\mu}h^{\mu\nu}\partial_{\nu}h^{\lambda}_{\;\;\lambda}\right\}
\nonumber \\
&&-\left(\partial^{\nu}h_{\nu\mu}-\frac{1}{2}\partial_{\mu}h^{\nu}
_{\;\;\nu}\right)\left(\partial^{\lambda}h_{\lambda}^{\mu}
-\frac{1}{2}\partial^{\mu}h^{\lambda}_{\;\;\lambda}\right)
\end{eqnarray}
We obtain the field equa\-tion
\begin{equation}\label{eq3}
\Box h_{\mu\nu}-\frac{1}{2}\eta_{\mu\nu}\Box h^{\lambda}_{\;\;\lambda}=0
\end{equation}
The trace of (\ref{eq3}) is
\begin{equation}\label{eq4}
\frac{2-n}{2}\Box h^{\lambda}_{\;\;\lambda}=0 \qquad \Longrightarrow
\qquad \Box h^{\lambda}_{\;\;\lambda}=0 \quad (n > 2)
\end{equation}
Then
\begin{equation}\label{eq5}
\Box h_{\mu\nu}=0
\end{equation}
The conjugate momentum is
\begin{eqnarray}
\pi^{\mu\nu}&=&\partial_{0}h^{\mu\nu}-\eta^{\mu\nu}
\left(\partial_{0}h^{\lambda}_{\;\;\lambda}
+\partial_{\lambda}h^{\lambda}_{\;\;0}+\frac{1}{2}B^{0}\right)
\nonumber \\
&& +\eta^{\mu 0}\left(\partial_{\lambda}h^{\lambda\nu}
-\frac{1}{2}\partial^{\nu}h^{\lambda}_{\;\;\lambda}+\frac{1}{2}B^{\nu}\right)
\nonumber \\
&&+\eta^{\nu 0}\left(\partial_{\lambda}h^{\lambda\mu}
-\frac{1}{2}\partial^{\mu}h^{\lambda}_{\;\;\lambda}+\frac{1}{2}B^{\mu}\right)
\nonumber \\
&\approx&\partial_{0}h^{\mu\nu}-\frac{1}{2}\eta^{\mu\nu}\partial_{0}
h^{\lambda}_{\;\;\lambda}
\end{eqnarray}
The $n$ dimensional commutation relations among the components of $h_{\mu\nu}$
at two arbitrary space-time points
\begin{eqnarray}
\left[h_{\mu\nu}(x), \;h_{\alpha\beta}(y)\right]&=&\frac{i}{2} \{
\eta_{\mu\alpha}\eta_{\nu\beta}+\eta_{\mu\beta}\eta_{\nu\alpha}
\nonumber \\
&& \qquad \quad
-\frac{2}{n-2}\eta_{\mu\nu}\eta_{\alpha\beta} \} D^{(n)}(x-y)
\nonumber
\end{eqnarray}
The equal-time commutation relation are
\begin{eqnarray}\label{eq7}
\left[h_{\mu\nu}(x), \;\pi^{\lambda\rho}(y)\right]_{eq}
&=&\frac{i}{2}\left(\delta^{\lambda}_{\mu}\delta^{\rho}_{\nu}
+\delta^{\rho}_{\mu}\delta^{\lambda}_{\nu}\right)\delta^{(n-1)}(x-y)
\nonumber \\
\\
\left[h_{\mu\nu}(x), \;h_{\lambda\rho}(y)\right]_{eq}
&=&\left[\pi^{\mu\nu}(x), \;\pi^{\lambda\rho}(y)\right]_{eq}=0
\nonumber
\end{eqnarray}
The gra\-vi\-ton's propagator is
\[
\langle 0|T(h_{\mu\nu}(x)h_{\lambda\rho}(y)|0\rangle =-\frac{i}{2(2\pi)^{n}}
\int d^{n}k\;e^{ik(x-y)}\frac{X_{\mu\nu,\lambda\rho}}{k^{2}-i\epsilon}
\]
where
\[
X_{\mu\nu,\lambda\rho}=\eta_{\mu\lambda}\eta_{\nu\rho}
+\eta_{\mu\rho}\eta_{\nu\lambda}-\frac{2}{n-2}\eta_{\mu\nu}\eta_{\lambda\rho}
\]
The $S$-matrix element of the one gra\-vi\-ton exchange diagram is
\begin{widetext}
\begin{equation}
S_n=\frac{4\pi iG_{n}}{(2\pi)^{n-2}}
\left(p^{0}_{1}p^{0}_{2}q^{0}_{1}q^{0}_{2}\right)^{-1/2}
\left[p^{\mu}_{1}q^{\nu}_{1}-\frac{1}{2}\eta^{\mu\nu}
(p_{1}\cdot q_{1}+m^{2}_{1})\right]
\left[p^{\alpha}_{2}q^{\beta}_{2}-\frac{1}{2}\eta^{\alpha\beta}
(p_{2}\cdot q_{2}+m^{2}_{2})\right]
\times \frac{X_{\mu\nu,\alpha\beta}}{k^{2}}
\delta^{(n)}(p_{1}+p_{2}-q_{1}-q_{2})
\label{eq:So}
\end{equation}
\end{widetext}
which in turn yields the $T$-matrix element
\begin{equation}
T_n =-\frac{4G_{n}}{(2\pi)^{n-2}}\cdot \frac{n-3}{n-2}\cdot
\frac{m_{1}m_{2}}{k^{2}}
\end{equation}
in the static approximation in $n$ di\-men\-sions.
The reader should not be fooled by the apparent sameness of $S_n$ from GRT
in Eq.~(\ref{eq:So}) and the $S_n$ of dilatonic gravity in Eq.~(\ref{eq:Sd}): the
$X_{\mu\nu,\lambda\rho}$ matrix elements are distinct!
Thus for $n=3$, $T_{3}({\bf k})=0$
in the static approximation. Then there exists no static potential in the
lowest order in $2+1$-dimensional Einstein gravity.
The potentials in $n=3, 4$ are
\begin{equation}
V_3=0 \qquad V_4 =-\frac{G_{4}m_{1}m_{2}}{r} ~.
\end{equation}
The potential $V_4$ is in agreement with (\ref{pot4}). However the potential for $n=3$
vanishes, and the potential for $n=2$ diverges.  This latter
situation can be dealt with by setting $G_{n}=(1-\frac{n}{2})G_{2}$
and taking the $n\rightarrow 2$ limiting method of Mann and Ross \cite{2dross},
which yields the two-dimensional $T$-matrix element
\begin{equation}
T_2 =-\frac{2G_{2}m_{1}m_{2}}{k^{2}}
\end{equation}
and the potential is calculated from Eq.~(\ref{pot1}) as
\begin{equation}
V_2 =
-2G_{2} m_{1}m_{2}\int dk\frac{e^{-ikx}}{k^{2}}
=2\pi G_{2}m_{1}m_{2}r
\end{equation}
Note that for $n=2$, we cannot get the consistent
quantization of the theory based on the free-gra\-vi\-ton Lagrangian
derived from $\frac{2}{\kappa^{2}}\sqrt{-g}R$. For example, the
propagator can not be defined in the case of $n=2$.
In the dilaton theory the dilaton contributes to the Newtonian
potential ``indirectly'' through the mixing with the gra\-vi\-ton.

Unlike GRT, $3$-dimensional dilaton gravity ($2+1$) includes the
Newtonian potential in any dimension, once the gravitational constant
is appropriately rescaled. In this sense the theory of
gravity (\ref{Lag1}) we consider is a relativistic extension of
Newtonian gravity in $d+1$ di\-men\-sions. GRT, on the
other hand, does not include Newtonian gravity in $2+1$ di\-men\-sions
and is empty in $1+1$ di\-men\-sions. In the latter case, an appropriate
rescaling of Newton's constant yields the theory (\ref{Lag1}) in
the $n\to 2$ limit \cite{2dross}.
\section{Invariant function for massless field in $n$ dimensions}
\label{appb}
This section elaborates $D^{(n)}$ which is used
in appendix \ref{appa}.  Consider the following massless field,
\[
\pha(x)=\frac{1}{\sqrt{2}(2\pi)^{(n-1)/2}} \int_{k_{0}>0}
\frac{d^{n-1}{\bf k}}{k_{0}} \left( e^{ikx}a_{k}+e^{-ikx} a^{\dagger}_{k}
\right)
\]
\[
\left[a_{{\bf k}}, \;a^{\dagger}_{{\bf k}^{\prime}}\right]=k_{0}
{\delta}^{n-1} ({\bf k-k^{\prime}})
\]
\begin{eqnarray}
\left[\pha(x), \;\pha(y)\right]&=&\frac{1}{2(2\pi)^{n-1}}\int_{k_{0}>0}
\frac{d^{n-1}k}{k_{0}}\left\{e^{ik(x-y)}-e^{-ik(x-y)}\right\}
\nonumber \\
&=&iD^{(n)}(x-y)
\nonumber
\end{eqnarray}
\begin{eqnarray}
D^{(n)}(x)&=&-\frac{i}{2(2\pi)^{n-1}}\int_{k_{0}>0}
\frac{d^{n-1}k}{k_{0}}\left\{e^{ikx}-e^{-ikx}\right\}
\nonumber \\
&=&-\frac{1}{(2\pi)^{n-1}}\int_{k_{0}>0}\frac{d^{n-1}k}{k_{0}}
e^{i{\bf kx}}\mbox{sin}k_{0}x_{0}
\nonumber \\
&=&-\frac{i}{(2\pi)^{n-1}}\int d^{n}k\;\epsilon(k_{0})\delta(k^{2})e^{ikx}
\nonumber
\end{eqnarray}
\begin{eqnarray}
D^{(n)}(-x)&=&-D^{(n)}(x)
\nonumber \\
D^{(n)}(\Lambda x)&=&D^{(n)}(x)
\nonumber \\
\Box D^{(n)}(x)&=&0
\nonumber \\
D^{(n)}({\bf x},\;0)&=&0
\nonumber \\
\left.\frac{\partial}{\partial x_{0}}D^{(n)}(x)\right|_{x_{0}=0}
&=&-\delta^{(n-1)}({\bf x})
\nonumber
\end{eqnarray}

\section{Setting of metric terms $g_{\mu 0}$ }
\label{appc}


Concerning the setting the lapse and shift functions,
DeWitt wrote\cite{dewitt}:
\begin{quote}
 ``If desired, one can always assign definite va\-lues to $N$ and $N_a$ ($\alpha$ and $\beta_a$
in DeWitt's notations) which may be purely numerical or may depend on the $\gamma_{ab}$
 and $\pi^{ab}$. Each choice corresponds to the imposition of certain conditions on the
 space-time coordinates. For example, one may choose $N=1$, $N_a=0$ \ldots''
\end{quote}
(and consequently $N^{a} = g^{ab} N_b = 0$.)
Since the Wheeler-DeWitt Quantum Geometrodynamics is supposed to be gauge invariant,
a concrete choice of $N$, $N_a$, $N^a$ is believed {\em not} to have any significance.
York [P.~8 of \cite{york1}] points out:
\begin{quote}
``In this "canonical"-like $3+1$ form, there are no time derivatives of $N$  or of $N^a$ \ldots
we have an easy way of seeing that $\dot{N}$ and $\dot{N}^a$ are dynamically irrelevant''.
\end{quote}
However, this notion has been challenged to some extent by Tatyana Shes\-ta\-ko\-va \cite{shestakova1,shestakova2}
who advocates an ``extended phase space'' approach and insists that the choice of $N$, $N_a$,
is actually a choice of gauge conditions and affects the resulting physical picture. Natalia Kiriushcheva
criticized the work of Shes\-ta\-ko\-va \cite{natalia1,natalia2} and points out:
\begin{quote}
``Dirac made an additional assumption that $g_{0a} = 0$ which noticeably
simplified his calculations but also led him to the conclusion that ”this simplification
can be achieved only at the expense of abandoning four-dimensional symmetry”.\cite{dirac}''
\end{quote}
Yet, Kiriushcheva admits\cite{natalia1}:
\begin{quote}
``\ldots we show that his assumption $g_{0a} = 0$, used to simplify his
calculation of diffe\-rent con\-tri\-bu\-tions to the secondary constraints, is unwarranted; yet, remarkably
his total Hamiltonian is equivalent to the one computed without the assumption $g_{0a} = 0$.''
\end{quote}
Shes\-ta\-ko\-va responded to the criticism by Kiriushcheva and her collaborators\cite{shestakova3}.
In the opinion of Natalia Ki\-riush\-cheva, the ADM approach already contains the loss of physics
as this representation res\-tricts possible coordinate transformations: a space-like hypersurface remains space-like.
Ki\-riush\-cheva claims it contradicts the main principle of GRT, the principle of general co\-va\-riance
(according to which any possible transformation of a coordinate system should be acceptable).

However, the ADM formulation was used for the Hamiltonian analysis where, \eg a time coordinate is singled out and
different treatment of time and space coordinates does not lead to contradictions.
It proved invaluable in decoupling the field equa\-tions for the $1+1$ case.
If, in addition, the lapse and shift are set to constants,  Ki\-riush\-cheva believes this will produce further ``destruction of
physics'', specifically the general class of physics solutions,
because such an operation will mean a substitution of the $3+1$ picture of the world, which is the essence of General and Special Relativity,
by a $3$-dimensional description, \ie Newtonian mechanics.

Notwithstanding
the contention of either Ki\-riush\-cheva or Shes\-ta\-ko\-va with what is now conventional wisdom (and even
the disagreement between themselves), we take the point of view that Dirac was essentially correct.
The ADM approach and the assump\-tions $g_{0a} = 0$ do res\-trict the class of so\-lu\-tions ob\-tainable.  However,
we are not seeking \eg Kerr metric solutions with a manifest $4 \times 4 $ co\-va\-riance where the $g_{0 \nu}$
are nonzero and vary
in time.  For an interacting system of point-particles, these assumptions can potentially yield
a realistic class of solutions with departures from these assumptions addressed subsequently.

In practice, $g_{0a} = 0$ often serves as {\em initial conditions} with departures obtained from the
time evolution of the gi\-ven system, as is often the case in numerical relativity where the equa\-tions
determining $N$ and $N_a$ are obtained by taking the time derivatives of the coordinate conditions\cite{adm0}.
Note that departures from these starting assumptions for a matter Lagrangian of Eq.~(\ref{eq:LM}) (last term of Eq.~(\ref{eq:LM}))
in the absence of an external
magnetic field has been worked out by Ki\-mura with 
analy\-tical so\-lu\-tions [Eqs.~$(3.18)-(3.20)$ of  \cite{r7}].

These arguments were made with respect to GRT but also apply to our scalar-tensor theory in particular since the latter
becomes GRT in the limit $\Psi \rightarrow 1$ and deviations from GRT are small. Similarly, they also apply
to a range of matter Lagrangians (\eg individual terms of Eq.~(\ref{eq:LM}) and in totality).

\section{On the generator and coordinate conditions}
\label{appd}
In $3+1$ dimensional GRT, it is commonly known that one of the coordinate conditions
appropriate for particle dynamics is 
[Eqs.~$(4.22a)$ and $(4.22b)$ of \cite{adm0}]
\begin{eqnarray}
t&=&-\frac{1}{2\triangle}\left(\pi^{T}
+\frac{1}{\triangle}\pi^{cd}_{\;\;,cd}\right)=-\frac{1}{2\triangle}\pi^{aa} \label{N7-1} \\
x^{a}&=&h_{a}-\frac{1}{4\triangle}h^{T}_{\;,a} \label{N7-2}
\end{eqnarray}
where $\triangle = \nabla^2$ is the Laplacian in $3$-space.  The total generator is
\begin{equation}
G=G_M
+\int d^{3}x \pi^{ab} \delta g_{ab}
\label{eq:G}
\end{equation}
where $G_M$ refers to the part dependent on the matter Lagrangian; using
the orthogonal decomposition of Eqs.~(\ref{eq:decomp}), $G$ is transformed as:
\begin{widetext}
\begin{eqnarray}
G&=&
G_M
+\int d^{3}x  \bigg\{ {\pi^{abTT}}
\delta h^{TT}_{ab}
+ \left[\frac{1}{2}(\delta_{ab}
-\frac{1}{\triangle}\partial_{a}\partial_{b})\pi^{T}+\pi^{a}_{\;,b}
+\pi^{b}_{\;,a}\right]
\delta\left[\frac{1}{2}\delta_{ab} h^{T}
-\frac{1}{2\triangle}\partial_{a}\partial_{b}h^{T}+h_{a,b}+h_{b,a}\right]
\bigg\}
\nonumber \\
&=&
G_M
+\int d^{3}x
\bigg\{
\pi^{abTT}
\delta h^{TT}_{ab}
+(\frac{1}{2}\pi^{T}+\pi^{a}_{\;,a})\delta h^{T}
+(\pi^{ab}-\pi^{abTT})\delta\left[-\frac{1}{2\triangle}\partial_{a}\partial_{b}
h^{T}+h_{a,b}+h_{b,a}\right]\bigg\}
\nonumber \\
&=&
G_M
+\int d^{3}x
\bigg\{
\pi^{abTT} \delta h^{TT}_{ab}
- 
\triangle h^{T}\delta\left[\frac{1}{2\triangle}(\pi^{T}
+\frac{1}{\triangle}\pi^{ca}_{\;,ca})\right]-2\pi^{ab}_{\;,b}\delta\left[
h_{a}-\frac{1}{4\triangle}h^{T}_{\;,a}\right]\bigg\} \nonumber
\end{eqnarray}
\end{widetext}
In this transformation the surface terms are discarded. Of course the
vanishing of the surface terms has been checked. (Variations at spatial
infinity are consistently set to zero.)
From this expression we set the coordinate condition Eqs.~(\ref{N7-1}) and (\ref{N7-2}).
The differential form of the conditions of (\ref{N7-1}) and (\ref{N7-2}) are
[Eqs~$(4.22c)$ and $(4.22d)$ of \cite{adm0}]
\begin{eqnarray}
\triangle g_{ab,b}-\frac{1}{4} g_{bc,abc}-\frac{1}{4} \triangle g_{bb,a}& =& 0
\\
\pi^{aa}= {\pi^{aa}}_{GRT} &= &0 ~.
\end{eqnarray}
The solution of the metric tensor is [Eq.~$(2.12)$ of \cite{r7} and
Eq.~$(2.19)$ of \cite{r8}]
\begin{equation}
\gamma_{ab} = g_{ab} =  \delta_{ab} ( 1 + \frac{1}{2} h^{T}  ) +  h_{ab}^{TT}
\end{equation}

\section{Divergences of Eq. ~(\ref{eq:pidiff})}
\label{appe}

Holding the Taub function $\alpha = N/\sqrt{h}$ of Eq.~(\ref{eq:taub}) constant and expanding 
Eq.~(\ref{eq:pidiff}) 
and  excluding the term in $\pi$, we obtain
\begin{equation}
\Pi \partial_a \Psi ~ \rightarrow ~
 \frac{3}{2} \underbrace{\frac{\partial_t \Psi \partial_a \Psi  }{\Psi}}_{(1)}
 - \frac{1}{2} \underbrace{\partial_t \Psi \partial_a \Psi }_{(2)}
\label{eq:48}
\end{equation}
Substituting $\Psi(t,x,y,z) = G(t) F(x,y,z) $ into the first 
part \ie{} the term $(1)$ of Eq.~(\ref{eq:48}) yields
\[
\frac{\partial_t \Psi \partial_a \Psi  }{\Psi}
~ = ~ \frac{d G(t) }{dt} \frac{d F(x,y,z) }{d x_a }
\]
This is clearly a divergence. For example, set $x_a=x$
\begin{eqnarray}
\int d^3 x ~ \frac{d G(t) }{dt} \frac{d F(x,y,z) }{d x }& = & \frac{d G(t) }{dt} \int dy dz \left. F(x,y,z) \right]_{-\infty}^{\infty} \nonumber \\
       & = & 0 \nonumber
\end{eqnarray}
for any function of the form $F = 1+\phi(x,y,z)$, as in Eq.~(\ref{dila}) where $\psi$ vanishes at infinity, and similarly for $x_a = y$ and $z$.
In view of the functional form of Eq.~(\ref{dila}), consider now a sum \ie{}
$\Psi(t,x,y,z)  =  G(t) +  F(x,y,z) $, into the $(1)$ term of Eq.~(\ref{eq:48})
\begin{eqnarray}
\frac{\partial_t \Psi \partial_a \Psi  }{\Psi}
& =  & \frac{d G(t) }{dt} \frac{d F(x,y,z) }{d x_a }/\left[G(t) + F(x,y,z) \right] \nonumber \\
&= & \frac{d G(t) }{dt} \frac{d}{d x_a} \ln ( G(t) + F(x,y,z) ) \nonumber
\end{eqnarray}

Here mixed terms in $t$ and spa\-tial coordinates appear.  However, $ G(t)$ and $G'(t)$ 
depend only on $t$ and are
therefore constant relative to the integration over spa\-tial coordinates.  Again we have a divergence for the $d/d x_a$ term.
 
We repeat this exercise for term $(2)$ of Eq.~(\ref{eq:48}) with 
$\Psi(t,x,y,z)  =  G(t) F(x,y,z)$
\begin{eqnarray}
\partial_t \Psi \partial_a \Psi  & = & 
\frac{d G(t) }{dt} G(t) \frac{d F(x,y,z) }{d x_a }  F(x,y,z)  \nonumber \\
&= & \frac{1}{4} \frac{d (G(t)^2 ) }{dt} \frac{d (F(x,y,z)^2 ) }{d x_a} \nonumber
\end{eqnarray}
This is again a divergence similar to the second case of $(1)$.
Finally,
for $\Psi(t,x,y,z)  =  G(t) +  F(x,y,z)$, we have
\[
\partial_t \Psi \partial_a \Psi  
~ = ~  \frac{d G(t) }{dt} \frac{d F(x,y,z) }{d x_a } 
\]
which is also clearly a divergence just like in the first case of $(1)$.

The term of Eq.~(\ref{eq:pidiff}), 
with $\pi$ gi\-ven by Eq.~(\ref{eq:piiso}) for isotropic coordinates, is
\begin{equation}
\pi ~ = ~ - \frac{3}{2} \sqrt{ \gamma } \dot{\Psi} 
\partial_a [ \ln (\Psi ) ]
\end{equation}
The Taub condition with the simplifications of Eq.~(\ref{simp}), 
\ie{} $N=1$,
implies that $\gamma = \mbox{const.} $,
and therefore the term in square brackets is clearly a divergence.

\section{Derivation of Eq.~(\ref{eq:jose0})}
\label{appf}

$\Gamma^{a}_{nb} $ is the Christoffel symbol of the second kind,
\[
\Gamma^{a}_{nc} ~ = ~ \gamma^{am} \Gamma_{mnc} = \frac{1}{2}  \gamma^{am} ( \gamma_{mn,c}
+ \gamma_{mc,n} - \gamma_{nc,m} )
\]
and is symmetric in the two lower indices.
We have
\begin{eqnarray}
D_a ( \pi^a_b ) =  \partial_a \pi^a_b  & + & 
\Gamma_{ac}^a \pi_b^c - \Gamma_{ab}^c \pi_c^a \nonumber \\
  =  \partial_a \pi^a_b 
& + & \frac{1}{2} \gamma^{ae} \left(
\partial_a \gamma_{ec} + \partial_c \gamma_{ea} - \partial_e \gamma_{ac} 
\right) \pi_b^c \nonumber \\
& - & \frac{1}{2} \gamma^{ce} \left(
\partial_a \gamma_{eb} + \partial_b \gamma_{ae} - \partial_e \gamma_{ab} 
\right) \pi_c^a \nonumber \\
  =  \partial_a \pi^a_b 
& + & \frac{1}{2} \gamma^{ae} \left(
\underbrace{\partial_a \gamma_{ec} - \partial_e \gamma_{ac}}_{0} + \partial_c \gamma_{ea} 
\right) \pi_b^c \nonumber \\
& - & \frac{1}{2} \left(
\underbrace{\partial_a \gamma_{eb} - \partial_e \gamma_{ab}}_{0} + \partial_b \gamma_{ae} 
\right) \pi^{ea} \nonumber \\
  =  \partial_a \pi^a_b 
& - & \frac{1}{2} \pi^{ae} \partial_b \gamma_{ae} + \frac{1}{2}
\underbrace{\gamma^{ae} \partial_c \gamma_{ea}}_{Tr ( \partial_c \ln (\gamma)) }
\pi_b^c  \label{eq:jose}
\end{eqnarray}
where we have used the symmetry of $\gamma_{ab}$ and $\pi_{ab}$. 
Equation (\ref{eq:jose}) is the result of Eq.~(\ref{eq:jose0}) 
with $a$ and $b$ interchanged.
The last term in Eq.~(\ref{eq:jose}) involves the logarithmic derivative of the
determinant of the metric.  It goes to zero if $\gamma=1$ or if the volume (whose
element is proportional to this term) is {\em fixed} within ADM.  Although
this is the case for many applications of ADM, this could be in doubt in
\eg{}, cosmological studies of an expanding universe.  However, it can be
justified if, for example,  the Taub function $\alpha$ of Eq.~(\ref{eq:taub}) is unit
or a constant.




\begin{thebibliography}{49}%
\makeatletter
\providecommand \@ifxundefined [1]{%
 \@ifx{#1\undefined}
}%
\providecommand \@ifnum [1]{%
 \ifnum #1\expandafter \@firstoftwo
 \else \expandafter \@secondoftwo
 \fi
}%
\providecommand \@ifx [1]{%
 \ifx #1\expandafter \@firstoftwo
 \else \expandafter \@secondoftwo
 \fi
}%
\providecommand \natexlab [1]{#1}%
\providecommand \enquote  [1]{``#1''}%
\providecommand \bibnamefont  [1]{#1}%
\providecommand \bibfnamefont [1]{#1}%
\providecommand \citenamefont [1]{#1}%
\providecommand \href@noop [0]{\@secondoftwo}%
\providecommand \href [0]{\begingroup \@sanitize@url \@href}%
\providecommand \@href[1]{\@@startlink{#1}\@@href}%
\providecommand \@@href[1]{\endgroup#1\@@endlink}%
\providecommand \@sanitize@url [0]{\catcode `\\12\catcode `\$12\catcode
  `\&12\catcode `\#12\catcode `\^12\catcode `\_12\catcode `\%12\relax}%
\providecommand \@@startlink[1]{}%
\providecommand \@@endlink[0]{}%
\providecommand \url  [0]{\begingroup\@sanitize@url \@url }%
\providecommand \@url [1]{\endgroup\@href {#1}{\urlprefix }}%
\providecommand \urlprefix  [0]{URL }%
\providecommand \Eprint [0]{\href }%
\providecommand \doibase [0]{http://dx.doi.org/}%
\providecommand \selectlanguage [0]{\@gobble}%
\providecommand \bibinfo  [0]{\@secondoftwo}%
\providecommand \bibfield  [0]{\@secondoftwo}%
\providecommand \translation [1]{[#1]}%
\providecommand \BibitemOpen [0]{}%
\providecommand \bibitemStop [0]{}%
\providecommand \bibitemNoStop [0]{.\EOS\space}%
\providecommand \EOS [0]{\spacefactor3000\relax}%
\providecommand \BibitemShut  [1]{\csname bibitem#1\endcsname}%
\let\auto@bib@innerbib\@empty
\bibitem [{\citenamefont {Mann}\ and\ \citenamefont
  {Ohta}(1997)}]{mann:gravity1}%
  \BibitemOpen
  \bibfield  {author} {\bibinfo {author} {\bibfnamefont {R.~B.}\ \bibnamefont
  {Mann}}\ and\ \bibinfo {author} {\bibfnamefont {T.}~\bibnamefont {Ohta}},\
  }\href@noop {} {\bibfield  {journal} {\bibinfo  {journal} {Phys. Rev. D}\
  }\textbf {\bibinfo {volume} {55}},\ \bibinfo {pages} {4723} (\bibinfo {year}
  {1997})}\BibitemShut {NoStop}%
\bibitem [{\citenamefont {Mann}\ and\ \citenamefont {Ross}(1993)}]{2dross}%
  \BibitemOpen
  \bibfield  {author} {\bibinfo {author} {\bibfnamefont {R.~B.}\ \bibnamefont
  {Mann}}\ and\ \bibinfo {author} {\bibfnamefont {S.~F.}\ \bibnamefont
  {Ross}},\ }\href@noop {} {\bibfield  {journal} {\bibinfo  {journal} {Class.
  Quantum Grav.}\ }\textbf {\bibinfo {volume} {10}},\ \bibinfo {pages} {1405}
  (\bibinfo {year} {1993})}\BibitemShut {NoStop}%
\bibitem [{\citenamefont {Ohta}\ and\ \citenamefont
  {Mann}(1996)}]{mann:gravity0}%
  \BibitemOpen
  \bibfield  {author} {\bibinfo {author} {\bibfnamefont {T.}~\bibnamefont
  {Ohta}}\ and\ \bibinfo {author} {\bibfnamefont {R.~B.}\ \bibnamefont
  {Mann}},\ }\href@noop {} {\bibfield  {journal} {\bibinfo  {journal} {Class.
  Quantum Grav.}\ }\textbf {\bibinfo {volume} {13}},\ \bibinfo {pages} {2585}
  (\bibinfo {year} {1996})}\BibitemShut {NoStop}%
\bibitem [{\citenamefont {Farrugia}\ \emph {et~al.}(2007)\citenamefont
  {Farrugia}, \citenamefont {Mann},\ and\ \citenamefont
  {Scott}}]{mann:gravity2}%
  \BibitemOpen
  \bibfield  {author} {\bibinfo {author} {\bibfnamefont {P.~S.}\ \bibnamefont
  {Farrugia}}, \bibinfo {author} {\bibfnamefont {R.~B.}\ \bibnamefont {Mann}},
  \ and\ \bibinfo {author} {\bibfnamefont {T.~C.}\ \bibnamefont {Scott}},\
  }\href@noop {} {\bibfield  {journal} {\bibinfo  {journal} {Class. Quantum
  Grav.}\ }\textbf {\bibinfo {volume} {24}},\ \bibinfo {pages} {4647} (\bibinfo
  {year} {2007})}\BibitemShut {NoStop}%
\bibitem [{\citenamefont {Mann}\ and\ \citenamefont
  {Chak}(2002)}]{Mann:2001fg}%
  \BibitemOpen
  R.~B.~Mann and P.~Chak, Phys. Rev. {\bf E65}, 026128 (2002),
  arXiv:gr-qc/0101106 [gr-qc].
\bibitem [{\citenamefont {Burnell}\ \emph {et~al.}(2003)\citenamefont
  {Burnell}, \citenamefont {Mann},\ and\ \citenamefont
  {Ohta}}]{Burnell:2002ps}%
  \BibitemOpen
  F.~J.~ Burnell, R.~B.~Mann, and T.~Ohta,
  Phys. Rev. Lett. {\bf 90},
  134101 (2003),
  arXiv:gr-qc/0208044 [gr-qc].
\bibitem [{\citenamefont {Burnell}\ \emph {et~al.}(2004)\citenamefont
  {Burnell}, \citenamefont {Malecki}, \citenamefont {Mann},\ and\ \citenamefont
  {Ohta}}]{Burnell:2003iu}%
  \BibitemOpen
  F.~J.~Burnell, J.~J.~Malecki, R.~B.~Mann, and T.~Ohta,
  Phys. Rev. {\bf E69}
  016214 (2004),
  arXiv:gr-qc/0301099 [gr-qc].
\bibitem [{\citenamefont {Sikkema}\ and\ \citenamefont
  {Mann}(1991)}]{mann:gravity3}%
  \BibitemOpen
  \bibfield  {author} {\bibinfo {author} {\bibfnamefont {A.~E.}\ \bibnamefont
  {Sikkema}}\ and\ \bibinfo {author} {\bibfnamefont {R.~B.}\ \bibnamefont
  {Mann}},\ }\href@noop {} {\bibfield  {journal} {\bibinfo  {journal} {Class.
  Quantum Grav.}\ }\textbf {\bibinfo {volume} {8}},\ \bibinfo {pages} {219}
  (\bibinfo {year} {1991})}\BibitemShut {NoStop}%
\bibitem [{\citenamefont {Jackiw}(1984)}]{jackiw1}%
  \BibitemOpen
  \bibfield  {author} {\bibinfo {author} {\bibfnamefont {R.}~\bibnamefont
  {Jackiw}},\ }in\ \href@noop {} {\emph {\bibinfo {booktitle} {Quantum Theory
  of Gravity: Essays in Honor of the 60th Birthday of Bryce S DeWitt}}},\
  \bibinfo {editor} {edited by\ \bibinfo {editor} {\bibfnamefont {S.~M.}\
  \bibnamefont {Christensen}}}\ (\bibinfo  {publisher} {Adam Hilger Ltd},\
  \bibinfo {address} {Bristol},\ \bibinfo {year} {1984})\ pp.\ \bibinfo {pages}
  {403--20}\BibitemShut {NoStop}%
\bibitem [{\citenamefont {Jackiw}(1985)}]{jackiw2}%
  \BibitemOpen
  \bibfield  {author} {\bibinfo {author} {\bibfnamefont {R.}~\bibnamefont
  {Jackiw}},\ }\href@noop {} {\bibfield  {journal} {\bibinfo  {journal} {Nucl.
  Phys. B}\ }\textbf {\bibinfo {volume} {252}},\ \bibinfo {pages} {343}
  (\bibinfo {year} {1985})}\BibitemShut {NoStop}%
\bibitem [{\citenamefont {Teitelboim}(1984)}]{teitelboim}%
  \BibitemOpen
  \bibfield  {author} {\bibinfo {author} {\bibfnamefont {C.}~\bibnamefont
  {Teitelboim}},\ }in\ \href@noop {} {\emph {\bibinfo {booktitle} {Quantum
  Theory of Gravity: Essays in Honor of the 60th Birthday of Bryce S
  DeWitt}}},\ \bibinfo {editor} {edited by\ \bibinfo {editor} {\bibfnamefont
  {S.~M.}\ \bibnamefont {Christensen}}}\ (\bibinfo  {publisher} {Adam Hilger
  Ltd},\ \bibinfo {address} {Bristol},\ \bibinfo {year} {1984})\ p.\ \bibinfo
  {pages} {327}\BibitemShut {NoStop}%
\bibitem [{\citenamefont {Arnowitt}\ \emph
  {et~al.}(1962{\natexlab{a}})\citenamefont {Arnowitt}, \citenamefont {Deser},\
  and\ \citenamefont {Misner}}]{adm0}%
  \BibitemOpen
  \bibfield  {author} {\bibinfo {author} {\bibfnamefont {R.}~\bibnamefont
  {Arnowitt}}, \bibinfo {author} {\bibfnamefont {S.}~\bibnamefont {Deser}}, \
  and\ \bibinfo {author} {\bibfnamefont {C.}~\bibnamefont {Misner}},\
  }\href@noop {} {\emph {\bibinfo {title} {The Dynamics of General
  Relativity}}},\ edited by\ \bibinfo {editor} {\bibfnamefont {L.}~\bibnamefont
  {Witten}}\ (\bibinfo  {publisher} {Wiley},\ \bibinfo {address} {New York},\
  \bibinfo {year} {1962})\ pp.\ \bibinfo {pages} {227--265}
 \BibitemShut {NoStop}%
\bibitem [{\citenamefont {Kimura}(1961)}]{r7}%
  \BibitemOpen
  \bibfield  {author} {\bibinfo {author} {\bibfnamefont {T.}~\bibnamefont
  {Kimura}},\ }\href@noop {} {\bibfield  {journal} {\bibinfo  {journal} {Prog.
  Theor. Phys.}\ }\textbf {\bibinfo {volume} {26}},\ \bibinfo {pages} {157}
  (\bibinfo {year} {1961})}\BibitemShut {NoStop}%
\bibitem [{\citenamefont {Avery}(1995)}]{dimscal}%
  \BibitemOpen
  J.~Avery, in {\em Structure and Dynamics of Atoms and Mol\-e\-cules:
  Conceptual Trends}, edited by
  J.~L.~Calais and E.~S.~Kryachko (Springer, Netherlands, 1995) pp. 133--154.
\BibitemShut {NoStop}%
\bibitem [{\citenamefont {Scott}\ \emph
  {et~al.}(2006{\natexlab{a}})\citenamefont {Scott}, \citenamefont {Mann},\
  and\ \citenamefont {{\relax Martinez II}}}]{scott:aaecc}%
  \BibitemOpen
  \bibfield  {author} {\bibinfo {author} {\bibfnamefont {T.~C.}\ \bibnamefont
  {Scott}}, \bibinfo {author} {\bibfnamefont {R.~B.}\ \bibnamefont {Mann}}, \
  and\ \bibinfo {author} {\bibfnamefont {R.~E.}\ \bibnamefont {{\relax Martinez
  II}}},\ }\href@noop {} {\bibfield  {journal} {\bibinfo  {journal} {AAECC}\
  }\textbf {\bibinfo {volume} {17}},\ \bibinfo {pages} {41} (\bibinfo {year}
  {2006}{\natexlab{a}})}\BibitemShut {NoStop}%
\bibitem [{\citenamefont {Scott}\ \emph
  {et~al.}(2006{\natexlab{b}})\citenamefont {Scott}, \citenamefont {{\relax
  Aubert-Frecon}},\ and\ \citenamefont {Grotendorst}}]{scott:chemphys}%
  \BibitemOpen
  \bibfield  {author} {\bibinfo {author} {\bibfnamefont {T.~C.}\ \bibnamefont
  {Scott}}, \bibinfo {author} {\bibfnamefont {M.}~\bibnamefont {{\relax
  Aubert-Frecon}}}, \ and\ \bibinfo {author} {\bibfnamefont {J.}~\bibnamefont
  {Grotendorst}},\ }\href@noop {} {\bibfield  {journal} {\bibinfo  {journal}
  {Chem. Phys.}\ }\textbf {\bibinfo {volume} {324}},\ \bibinfo {pages} {323}
  (\bibinfo {year} {2006}{\natexlab{b}})}\BibitemShut {NoStop}%
\bibitem [{\citenamefont {Will}(2014)}]{Will:2014kxa}%
  \BibitemOpen
  C.~M.~Will, Living Rev. Rel. {\bf 17}, {4} (2014),
  arXiv:1403.7377 [gr-qc]
 \BibitemShut {NoStop}%
\bibitem [{\citenamefont {Bellazzini}\ \emph {et~al.}(2013)\citenamefont
  {Bellazzini}, \citenamefont {Csaki}, \citenamefont {Hubisz}, \citenamefont
  {Serra},\ and\ \citenamefont {Terning}}]{higgs}%
  \BibitemOpen
  \bibfield  {author} {\bibinfo {author} {\bibfnamefont {B.}~\bibnamefont
  {Bellazzini}}, \bibinfo {author} {\bibfnamefont {C.}~\bibnamefont {Csaki}},
  \bibinfo {author} {\bibfnamefont {J.}~\bibnamefont {Hubisz}}, \bibinfo
  {author} {\bibfnamefont {J.}~\bibnamefont {Serra}}, \ and\ \bibinfo {author}
  {\bibfnamefont {J.}~\bibnamefont {Terning}},\ }\href@noop {} {\bibfield
  {journal} {\bibinfo  {journal} {Eur. Phys. J. C}\ }\textbf {\bibinfo {volume}
  {73}},\ \bibinfo {pages} {2333} (\bibinfo {year} {2013})}\BibitemShut
  {NoStop}%
\bibitem [{\citenamefont {Zhang}\ and\ \citenamefont {Ma}(2011)}]{xiangdong1}%
  \BibitemOpen
  \bibfield  {author} {\bibinfo {author} {\bibfnamefont {X.}~\bibnamefont
  {Zhang}}\ and\ \bibinfo {author} {\bibfnamefont {Y.}~\bibnamefont {Ma}},\
  }\href {\doibase 10.1103/PhysRevD.84.104045} {\bibfield  {journal} {\bibinfo
  {journal} {Phys. Rev. D.}\ }\textbf {\bibinfo {volume} {84}},\ \bibinfo
  {pages} {104045} (\bibinfo {year} {2011})}\BibitemShut {NoStop}%
\bibitem [{\citenamefont {Arnowitt}\ \emph
  {et~al.}(1962{\natexlab{b}})\citenamefont {Arnowitt}, \citenamefont {Deser},\
  and\ \citenamefont {Misner}}]{adm1}%
  \BibitemOpen
  \bibfield  {author} {\bibinfo {author} {\bibfnamefont {R.}~\bibnamefont
  {Arnowitt}}, \bibinfo {author} {\bibfnamefont {S.}~\bibnamefont {Deser}}, \
  and\ \bibinfo {author} {\bibfnamefont {C.}~\bibnamefont {Misner}},\
  }\href@noop {} {\emph {\bibinfo {title} {Gravitation: An Introduction to
  Current Research}}}\ (\bibinfo  {publisher} {Wiley},\ \bibinfo {address} {New
  York},\ \bibinfo {year} {1962})\BibitemShut {NoStop}%
\bibitem [{\citenamefont {Arnowitt}\ \emph {et~al.}(1960)\citenamefont
  {Arnowitt}, \citenamefont {Deser},\ and\ \citenamefont {Misner}}]{adm2}%
  \BibitemOpen
  \bibfield  {author} {\bibinfo {author} {\bibfnamefont {R.}~\bibnamefont
  {Arnowitt}}, \bibinfo {author} {\bibfnamefont {S.}~\bibnamefont {Deser}}, \
  and\ \bibinfo {author} {\bibfnamefont {C.}~\bibnamefont {Misner}},\
  }\href@noop {} {\bibfield  {journal} {\bibinfo  {journal} {J. Math. Phys.}\
  }\textbf {\bibinfo {volume} {1}},\ \bibinfo {pages} {434} (\bibinfo {year}
  {1960})}\BibitemShut {NoStop}%
\bibitem [{\citenamefont {Misner}\ \emph {et~al.}(1973)\citenamefont {Misner},
  \citenamefont {Thorne},\ and\ \citenamefont {Wheeler}}]{gravitation}%
  \BibitemOpen
  \bibfield  {author} {\bibinfo {author} {\bibfnamefont {C.~W.}\ \bibnamefont
  {Misner}}, \bibinfo {author} {\bibfnamefont {K.~S.}\ \bibnamefont {Thorne}},
  \ and\ \bibinfo {author} {\bibfnamefont {J.~A.}\ \bibnamefont {Wheeler}},\
  }\href@noop {} {\emph {\bibinfo {title} {Gravitation}}}\ (\bibinfo
  {publisher} {W. H. Freeman},\ \bibinfo {address} {New York},\ \bibinfo {year}
  {1973})\BibitemShut {NoStop}%
\bibitem [{\citenamefont {DeWitt}(1967)}]{dewitt}%
  \BibitemOpen
  \bibfield  {author} {\bibinfo {author} {\bibfnamefont {B.}~\bibnamefont
  {DeWitt}},\ }\href@noop {} {\bibfield  {journal} {\bibinfo  {journal} {Phys.
  Rev.}\ }\textbf {\bibinfo {volume} {160}},\ \bibinfo {pages} {1113} (\bibinfo
  {year} {1967})}\BibitemShut {NoStop}%
\bibitem [{\citenamefont {Dirac}(1958)}]{dirac}%
  \BibitemOpen
  \bibfield  {author} {\bibinfo {author} {\bibfnamefont {P.~A.~M.}\
  \bibnamefont {Dirac}},\ }\href@noop {} {\bibfield  {journal} {\bibinfo
  {journal} {Proc. Roy. Soc.}\ }\textbf {\bibinfo {volume} {A246}},\ \bibinfo
  {pages} {333¨C343} (\bibinfo {year} {1958})}\BibitemShut {NoStop}%
\bibitem [{\citenamefont {Kiriushcheva}\ and\ \citenamefont
  {Kuzmin}(2011)}]{natalia1}%
  \BibitemOpen
  N.~Kiriushcheva and S.~V.~Kuzmin, 
  Central European Journal of Physics
  {\bf 9}, {576} (2011).
\BibitemShut {NoStop}%
\bibitem [{\citenamefont {Kiriushcheva}\ \emph {et~al.}(2011)\citenamefont
  {Kiriushcheva}, \citenamefont {Komorowski},\ and\ \citenamefont
  {Kuzmin}}]{natalia2}%
  \BibitemOpen
  N.~Kiriushcheva, P.~G.~Komorowski, and S.~V.~Kuzmin,
  ``Comment on "hamiltonian formulation for the theory of 
   gravity and canonical transformations in extended phase space'' 
   by T.~P.~Shestakova'' (2011), {http://arxiv.org/pdf/1107.2981.pdf}
\BibitemShut {NoStop}%
\bibitem [{\citenamefont {Ohta}\ \emph {et~al.}(1974)\citenamefont {Ohta},
  \citenamefont {Okamura}, \citenamefont {Kimura},\ and\ \citenamefont
  {Hiida}}]{r8}%
  \BibitemOpen
  \bibfield  {author} {\bibinfo {author} {\bibfnamefont {T.}~\bibnamefont
  {Ohta}}, \bibinfo {author} {\bibfnamefont {H.}~\bibnamefont {Okamura}},
  \bibinfo {author} {\bibfnamefont {T.}~\bibnamefont {Kimura}}, \ and\ \bibinfo
  {author} {\bibfnamefont {K.}~\bibnamefont {Hiida}},\ }\href@noop {}
  {\bibfield  {journal} {\bibinfo  {journal} {Prog. Theor. Phys.}\ }\textbf
  {\bibinfo {volume} {51}},\ \bibinfo {pages} {1598} (\bibinfo {year}
  {1974})}\BibitemShut {NoStop}%
\bibitem [{\citenamefont {Hamber}(2009)}]{hamber}%
  \BibitemOpen
  \bibfield  {author} {\bibinfo {author} {\bibfnamefont {H.~W.}\ \bibnamefont
  {Hamber}},\ }\href@noop {} {\emph {\bibinfo {title} {Quantum Gravitation The
  Feynman Path Integral Approach}}}\ (\bibinfo  {publisher} {Springer},\
  \bibinfo {address} {Berlin},\ \bibinfo {year} {2009})\BibitemShut {NoStop}%
\bibitem [{\citenamefont {{York}}(2006)}]{york1}%
  \BibitemOpen
  J.~W.~York, Jr., in
  {\em The Tenth Marcel Grossmann Meeting.  
  Proceedings of the MG10 Meeting held at Brazilian Center
  for Research in Physics (CBPF), Rio de Janeiro, Brazil, 20-26 July 2003,
  Eds.: M{\'a}rio Novello; Santiago Perez Bergliaffa; Remo Ruffini. Singapore:
  World Scientific Publishing, in 3 volumes, ISBN 981-256-667-8 (set), ISBN
  981-256-980-4 (Part A), ISBN 981-256-979-0 (Part B), ISBN 981-256-978-2 (Part
  C), 2006, XLVIII + 2492 pp.: 2006, p.3}, edited by
  M.~Novello, S.~ Perez Bergliaffa, and
  R.~Ruffini
  (2006) p.~3, {gr-qc/0405005} 
\BibitemShut {NoStop}%
\bibitem [{\citenamefont {Anderson}\ \emph {et~al.}(2000)\citenamefont
  {Anderson}, \citenamefont {Choquet-Bruhat},\ and\ \citenamefont {{\relax
  York, Jr.}}}]{taub1}%
  \BibitemOpen
  \bibfield  {author} {\bibinfo {author} {\bibfnamefont {A.}~\bibnamefont
  {Anderson}}, \bibinfo {author} {\bibfnamefont {Y.}~\bibnamefont
  {Choquet-Bruhat}}, \ and\ \bibinfo {author} {\bibfnamefont {J.~W.}\
  \bibnamefont {{\relax York, Jr.}}},\ }\href@noop {} {\bibfield  {journal}
  {\bibinfo  {journal} {Lect. Notes Phys.}\ }\textbf {\bibinfo {volume}
  {537}},\ \bibinfo {pages} {30} (\bibinfo {year} {2000})},\ \Eprint
  {http://arxiv.org/abs/gr-qc/9907099} {gr-qc/9907099} \BibitemShut {NoStop}%
\bibitem [{\citenamefont {Choquet-Bruhat}\ and\ \citenamefont {{\relax York,
  Jr.}}(2002)}]{taub2}%
  \BibitemOpen
  \bibfield  {author} {\bibinfo {author} {\bibfnamefont {Y.}~\bibnamefont
  {Choquet-Bruhat}}\ and\ \bibinfo {author} {\bibfnamefont {J.~W.}\
  \bibnamefont {{\relax York, Jr.}}},\ }\href@noop {} {\bibfield  {journal}
  {\bibinfo  {journal} {Lect. Notes Phys.}\ }\textbf {\bibinfo {volume}
  {592}},\ \bibinfo {pages} {29} (\bibinfo {year} {2002})},\ \Eprint
  {http://arxiv.org/abs/gr-qc/0202013} {gr-qc/0202013} \BibitemShut {NoStop}%
\bibitem [{\citenamefont {Jantzen}(2005)}]{taub3}%
  \BibitemOpen
  R.~T.~Jantzen, in
  {\em Elba Conference in Honor of Yvonne
  Choquet-Bruhat's 80th birthday: Analysis Manifolds and Geometric Structures 
  in Physics}, Vol.~119, edited by
  G.~Ferrarese (Bibliopolis, Naples, 2005) pp.~697--715,
  {gr-qc/0505086} 
\BibitemShut {NoStop}%
\bibitem [{\citenamefont {Sarbach}\ and\ \citenamefont
  {Tiglio}(2012)}]{Sarbach:2012pr}%
  \BibitemOpen
  \bibfield  {author} {\bibinfo {author} {\bibfnamefont {O.}~\bibnamefont
  {Sarbach}}\ and\ \bibinfo {author} {\bibfnamefont {M.}~\bibnamefont
  {Tiglio}},\ }\href {\doibase 10.12942/lrr-2012-9} {\bibfield  {journal}
  {\bibinfo  {journal} {Living Rev. Rel.}\ }\textbf {\bibinfo {volume} {15}},\
  \bibinfo {pages} {9} (\bibinfo {year} {2012})},\ \Eprint
  {http://arxiv.org/abs/1203.6443} {arXiv:1203.6443 [gr-qc]} \BibitemShut
  {NoStop}%
\bibitem [{\citenamefont {Calabrese}\ \emph
  {et~al.}(2002{\natexlab{a}})\citenamefont {Calabrese}, \citenamefont
  {Pullin}, \citenamefont {Sarbach},\ and\ \citenamefont
  {Tiglio}}]{Calabrese:2002ei}%
  \BibitemOpen
  G.~Calabrese, J.~Pullin, O.~Sarbach, and M.~Tiglio,
  Phys. Rev. {\bf D66}, 064011 (2002),
  {arXiv:gr-qc/0205073 [gr-qc]}
  \BibitemShut {NoStop}%
\bibitem [{\citenamefont {Calabrese}\ \emph
  {et~al.}(2002{\natexlab{b}})\citenamefont {Calabrese}, \citenamefont
  {Pullin}, \citenamefont {Sarbach},\ and\ \citenamefont
  {Tiglio}}]{Calabrese:2002ej}%
  \BibitemOpen
  G.~Calabrese, J.~Pullin, O.~Sarbach, and M.~Tiglio,
  Phys. Rev. {\bf D66}, 041501 (2002),
  {arXiv:gr-qc/0207018 [gr-qc]}
  \BibitemShut {NoStop}%
\bibitem [{Note1()}]{Note1}%
  \BibitemOpen
  \bibinfo {note} {There is a small typo in Kimura's paper\cite {r7} concerning
  the last term on the far right of his Eq.~ $(3.13)$ which should be $r_a$ not
  $r_{ab}$. This is rectified in his paper coauthored with T. Ohta\cite
  {r8}.}\BibitemShut {Stop}%
\bibitem [{\citenamefont {Avdeenkov}\ and\ \citenamefont
  {Zloshchastiev}(2011)}]{bose}%
  \BibitemOpen
  A.~V.~Avdeenkov and K.~Zloshchastiev,
  {J. Phys. B: At. Mol. Opt. Phys.}
  {\bf 44}, {195303} (2011), {arXiv:1108.0847}
  \BibitemShut {NoStop}%
\bibitem [{\citenamefont {Kartavenko}\ \emph {et~al.}(1998)\citenamefont
  {Kartavenko}, \citenamefont {Gridnev},\ and\ \citenamefont
  {Greiner}}]{nuclear}%
  \BibitemOpen
  V.~G.~Kartavenko, K.~A.~ Gridnev, and W.~ Greiner, {Int. J. Mod. Phys. E} 
 {\bf 7}, {287} (1998)
\BibitemShut {NoStop}%
\bibitem [{\citenamefont {Scott}\ \emph {et~al.}(2013)\citenamefont {Scott},
  \citenamefont {Fee},\ and\ \citenamefont {Grotendorst}}]{scott:sigsam1}%
  \BibitemOpen
  T.~C.~Scott, G.~J.~Fee, and J.~Grotendorst,
  SIGSAM {\bf 47}, {75} (2013)
\BibitemShut {NoStop}%
\bibitem [{\citenamefont {Scott}\ \emph {et~al.}(2014)\citenamefont {Scott},
  \citenamefont {Fee},\ and\ \citenamefont {Grotendorst}}]{scott:sigsam2}%
  \BibitemOpen
  T.~C.~Scott, G.~J.~Fee, and J.~Grotendorst,
  SIGSAM {\bf 48}, {42} (2014)
\BibitemShut {NoStop}%
\bibitem [{\citenamefont {Zloshchastiev}(2011)}]{svt}%
  \BibitemOpen
  \bibfield  {author} {\bibinfo {author} {\bibfnamefont {K.~G.}\ \bibnamefont
  {Zloshchastiev}},\ }\href@noop {} {\bibfield  {journal} {\bibinfo  {journal}
  {Acta Phys. Polon. B}\ }\textbf {\bibinfo {volume} {42}},\ \bibinfo {pages}
  {261} (\bibinfo {year} {2011})},\ \Eprint {http://arxiv.org/abs/0912.4139.}
  {arXiv:0912.4139.} \BibitemShut {NoStop}%
\bibitem [{\citenamefont {Stephani}(1990)}]{radiation}%
  \BibitemOpen
  \bibfield  {author} {\bibinfo {author} {\bibfnamefont {H.}~\bibnamefont
  {Stephani}},\ }\href@noop {} {\emph {\bibinfo {title} {General Relativity: An
  Introduction to the Theory of the Gravitational Field}}}\ (\bibinfo
  {publisher} {Cambridge University Press},\ \bibinfo {address} {Cambridge},\
  \bibinfo {year} {1990})\BibitemShut {NoStop}%
\bibitem [{\citenamefont {Feynman}(1995)}]{feynman}%
  \BibitemOpen
  \bibfield  {author} {\bibinfo {author} {\bibfnamefont {R.}~\bibnamefont
  {Feynman}},\ }\href@noop {} {\emph {\bibinfo {title} {The Feynman lectures on
  Gravitation}}},\ edited by\ \bibinfo {editor} {\bibfnamefont
  {B.}~\bibnamefont {Hatfield}}\ (\bibinfo  {publisher} {Addison-Wesley},\
  \bibinfo {address} {Reading, Mass.},\ \bibinfo {year} {1995})\BibitemShut
  {NoStop}%
\bibitem [{\citenamefont {Nambu}(1950{\natexlab{a}})}]{nambu1}%
  \BibitemOpen
  \bibfield  {author} {\bibinfo {author} {\bibfnamefont {Y.}~\bibnamefont
  {Nambu}},\ }\href@noop {} {\bibfield  {journal} {\bibinfo  {journal} {Prog.
  Theor. Phys.}\ }\textbf {\bibinfo {volume} {5}},\ \bibinfo {pages} {82}
  (\bibinfo {year} {1950}{\natexlab{a}})}\BibitemShut {NoStop}%
\bibitem [{\citenamefont {Nambu}(1950{\natexlab{b}})}]{nambu2}%
  \BibitemOpen
  \bibfield  {author} {\bibinfo {author} {\bibfnamefont {Y.}~\bibnamefont
  {Nambu}},\ }\href@noop {} {\bibfield  {journal} {\bibinfo  {journal} {Prog.
  Theor. Phys.}\ }\textbf {\bibinfo {volume} {5}},\ \bibinfo {pages} {321}
  (\bibinfo {year} {1950}{\natexlab{b}})}\BibitemShut {NoStop}%
\bibitem [{\citenamefont {Nambu}(1950{\natexlab{c}})}]{nambu3}%
  \BibitemOpen
  \bibfield  {author} {\bibinfo {author} {\bibfnamefont {Y.}~\bibnamefont
  {Nambu}},\ }\href@noop {} {\bibfield  {journal} {\bibinfo  {journal} {Prog.
  Theor. Phys.}\ }\textbf {\bibinfo {volume} {5}},\ \bibinfo {pages} {614}
  (\bibinfo {year} {1950}{\natexlab{c}})}\BibitemShut {NoStop}%
\bibitem [{\citenamefont {Shestakova}(2009)}]{shestakova1}%
  \BibitemOpen
 T.~P.~Shestakova, in
  {\em Physical Interpretations of Relativity Theory: Proceedings of International Scientific Meeting (Moscow, 6-9 July 2009)},
  edited by M.~C.~Duffy, V.~O.~Gladyshev, A.~N.~Morozov and P.~Rowlands (Moscow, 2009) p.~49
\BibitemShut {NoStop}%
\bibitem [{\citenamefont {Shestakova}(2011)}]{shestakova2}%
  \BibitemOpen
  \bibfield  {author} {\bibinfo {author} {\bibfnamefont {T.~P.}\ \bibnamefont
  {Shestakova}},\ }\href@noop {} {\bibfield  {journal} {\bibinfo  {journal}
  {Class. Quantum Grav.}\ }\textbf {\bibinfo {volume} {28}},\ \bibinfo {pages}
  {055009} (\bibinfo {year} {2011})}\BibitemShut {NoStop}%
\bibitem [{\citenamefont {Shestakova}(2014)}]{shestakova3}%
  \BibitemOpen
  \bibfield  {author} {\bibinfo {author} {\bibfnamefont {T.~P.}\ \bibnamefont
  {Shestakova}},\ }\href@noop {} {\bibfield  {journal} {\bibinfo  {journal}
  {Gravitation and Cosmology}\ }\textbf {\bibinfo {volume} {20}},\ \bibinfo
  {pages} {67} (\bibinfo {year} {2014})}\BibitemShut {NoStop}%
\end{thebibliography}

%
\end{document}